\documentclass[aps,reprint,amsmath,longbibliography]{revtex4-1} 

\bibpunct{}{}{,}{s}{}{}

\usepackage{dcolumn} 
\usepackage{graphicx} 
 \usepackage{braket}
 \usepackage{framed}

 \newcommand{\figwide}{7.8cm}

\usepackage{array} 
\newcolumntype{K}[1]{>{\centering\arraybackslash}p{#1}} 
\usepackage{longtable}

\newenvironment{dtable}[1]{
\begin{table*}[hbt]
\caption{#1}
\begin{ruledtabular}
\begin{tabular}[t]{K{.1\textwidth} p{.8\textwidth} K{.1\textwidth}}
No.&Name&Ref.\\
\hline
}{%
\end{tabular}
\end{ruledtabular}
\end{table*}
}

\usepackage{color}					% For editing purposes

\begin{document}

\title{Framing difficulties in quantum mechanics} 
\date{\today}

\author{Bahar Modir}
\author{John D. Thompson}
\author{Eleanor C. Sayre} 
\email{esayre@ksu.edu}
\affiliation{Department of Physics, Kansas State University, Manhattan, Kansas 66506}

\begin{abstract} 
Students' difficulties in quantum mechanics may be the result of unproductive framing rather than a fundamental inability to solve the problems or misconceptions about physics content. Using the theoretical lens of epistemological framing, we applied previously developed frames to seek an underlying structure to the long lists of published difficulties that span many topics in quantum mechanics. Mapping descriptions of published difficulties into errors in epistemological framing and resource use, we analyzed descriptions of students' problem solving to find their frames, and compared students' framing to the framing (and frame shifting) required by problem statements. We found three categories of error: mismatches between students' framing and problem statement framing; inappropriate or absent shifting between frames; and insufficient resource activation within an appropriate frame.
%Given this framework, we can predict the kinds of difficulties that will emerge for a given problem in quantum mechanics, yielding a possible deeper structure to student difficulties.

\end{abstract}


\maketitle

%%%%%%%%%%%%%%%%%%%%%%%%%%%%%%%%%%%%%%%%%%%% 
%% MAINMATTER 
%%%%%%%%%%%%%%%%%%%%%%%%%%%%%%%%%%%%%%%%%%%%

\section{Introduction}

Researchers in student understanding of quantum mechanics have used a ``difficulties" framework to understand student  reasoning, %(e.g. \cite{Singh2015,Emigh2015,Passante2015})
identifying long lists of difficulties which span many topics in quantum mechanics. 
The goal of research in quantum difficulties is to determine common, repeatable incorrect patterns of students' reasoning\cite{Styer1996, Singh2015, Emigh2015, Passante2015}. Researchers refer to identified difficulties as universal patterns, since they occur across a wide range of student populations despite varying academic backgrounds\cite{Singh2006}.

Although the realms of quantum and classical mechanics are different -- the classical world is simpler and more intuitive than the quantum world -- researchers have long considered the possibility of difficulties in quantum mechanics being analogous to misconceptions in classical mechanics\cite{Clement1982}.  This similarity is due to both  persistent misconceptions or difficulties in students' reasoning\cite{Singh2001}, and students not having enough preparation with the formalism of quantum mechanics\cite{Marshman2015}.

Research has detailed lists of student difficulties in determining the time dependency of stationary, superposed, and degenerate eigenfunctions\cite{Emigh2015}; the effect of time dependency of different physical systems on the probability densities\cite{Emigh2015}; energy measurements of a quantum mechanical system\cite{Passante2015}; concepts of the time-dependent Schr{\"o}dinger equation\cite{Passante2015} (TDSE); and the role of Hamiltonian physics in determining energy.\cite{Passante2015}  As additional research in student difficulties investigates other topics in QM, we expect that many additional difficulties will be found.  

However, we posit that these disparate difficulties can be unified through the lens of epistemological framing\cite{Tannen1993}, errors in frame transitions \cite{Irving2013framing}, and errors in the content of a frame (e.g. with the Resources Framework\cite{Hammer2000}). This paper presents a secondary analysis of published difficulties in quantum mechanics through the lens of epistemological framing. 

Our goal in this paper is to reanalyze students' difficulties in quantum mechanics. We apply a set of frames previously developed by our research team\cite{Modir2017QuantumFraming,Nguyen2016StuFraming} to a long list of published difficulties in quantum mechanics in order to find an underlying structure to them.  After developing our theoretical lens on our own video-based data, we turned to the published literature on student difficulties in quantum mechanics to seek an underlying structure to students' difficulties.  %Because this work is fundamentally about reinterpreting data analyzed under one theoretical framework to another framework, the details of that reinterpretation are quite dense.  

% In the remaining sections, we outline both theories (Sections \ref{sec:difficulties} and \ref{sec:framing}), our methods of secondary analysis (Section \ref{sec:methods}), and group difficulties by the kind of framing errors they represent (Sections \ref{sec:transition} through \ref{

%\section{Student understanding of Quantum Mechanics}

The choice of different theoretical frameworks is consequential for the kinds of data we collect, the way we analyze them, and the implications of our research for other researchers and for instructors.  Difficulties focuses our attention on the ways in which students' wrong answers prevent them from succeeding in problem solving; framing focuses our attention on the pathways that students take as they navigate a problem.  This difference of attention means that research using the two frameworks values two different kinds of evidence.  Difficulties research values survey responses -- possibly multiple choice -- to carefully crafted questions that elicit known patterns of wrong answers.  Framing research values records of students working on longer problems -- possibly in groups -- to see how their frames change in the course of the problem.  As more topics are researched, the list of difficulties grows enormous; however, the list of frames can remain small.  This feature suggests that framing, as a more parsimonious theory, can be more helpful to researchers seeking mechanisms for student understanding and instructors seeking to facilitate student reasoning during problem solving.  

\section{Theoretical framework}

\subsection{Difficulties}
%Due to the aforementioned similarities between the nature of difficulties in classical mechanics and quantum mechanics in quantum difficulties literature\cite{Singh2001, Marshman2015}, an explanation of the misconception perspective and its alternative view in classical mechanics would be helpful.

In a misconceptions or difficulties view, students apply an incorrect model of a concept across a wide range of situations independent of the context\cite{Clement1982, McCloskey1983}. The core of conceptual understanding occurs by confronting the incorrect conception, and replacing it with a new concept. This unitary view of students' reasoning guides our attention as researchers toward the identification of topics with which students have difficulties at the cost of missing students' epistemological changes\cite{Hammer1996morethan} because a difficulties view predicts a stable model of thinking that is repeatable, and does not account for sudden or contextual changes in the nature of student reasoning.

A large number of students showing the same wrong answer to the same question implies a widespread difficulty in a certain topic; if the same difficulty presents across multiple questions or over time, it is robust.
There have been many difficulties identified in quantum mechanics over the last 20 years across many different topics. 
 For example, there are several sub-topic difficulties reported related to the topic of time dependence of the wave function: incorrect belief that the time evolution of a wave function is always via an overall phase factor of the type $e^\frac{-iEt}{\hbar}$; inability to differentiate between $e^\frac{-iHt}{\hbar}$ and $e^\frac{-iEt}{\hbar}$; and belief that for a time-independent Hamiltonian, the wave function does not depend on time.\cite{Singh2015} 
 
Research into student difficulties is often focused on eliciting those difficulties in regular ways (possibly also involving the development of research-based conceptual assessments\cite{Madsen2017RBAI}), developing curricula to ameliorate those difficulties, and iteratively improving the curricula.  Common methods for the identification and documentation of difficulties are outlined in section \ref{sec:difficultiesmethods}; this paper is not concerned with the curriculum development or evaluation aspects of difficulties research.

Because difficulties research seeks to elicit regular, repeatable, wrong answers, much work on identifying difficulties relies on developing carefully phrased questions to elicit difficulties.  This development effort, while deeply important to work using this framework, is often omitted or given short shrift in difficulties research publications.  That omission is a consequence of the epistemic commitments of the framework: the framework assumes that student responses betray underlying and robust difficulties, and therefore presentations of those difficulties focus on students' ideas as more-or-less independent of the questions posed.  However, any student response is an interaction effect between the students' ideas and the question posed (as well as other contextual features such as question format and student identity).  To properly consider which student ideas are elicited in a response, it is necessary to examine the interaction with the question posed.

\subsection{Manifold views}
An alternate view to a unitary difficulties view is a manifold ``knowledge in pieces'' view.  In this view of student reasoning, we conceptualize student thinking as being highly context dependent and composed of small, reusable elements of knowledge and reasoning called ``pieces''. These pieces are not themselves correct or incorrect, but the ways in which students put them together to solve problems may be. By focusing on the pieces of student reasoning and how they fit together, this view of student reasoning foregrounds the seeds of productive student reasoning and not just incorrect answers. Theories in this family include phenomenological primitives\cite{diSessa1993} (p-prims), resources\cite{Hammer2000}, and symbolic forms\cite{Sherin2001}.

A strong thread of research using knowledge in pieces is to investigate students' epistemologies. Epistemological resources\cite{Hammer2003Tapping} connect to conceptual\cite{Hammer2000} and procedural\cite{Black2009Procedural} resources in networks\cite{Wittmann2006Conceptual} to help students solve problems. 

The mechanism that allows control of which subset of resources are activated locally in a given context is epistemological framing\cite{Redish2004}. Framing shows the nature of students' knowledge that emerges from a coherent set of fine-grained resources which coherently and locally work together in a situation\cite{Hammer2004rft}.  

Epistemological frames reveal students'\cite{Redish2004, Tannen1993} ways of thinking and expectations. They govern which ideas students link together and utilize to solve problems. Students' epistemological framing is highly context sensitive. Being in the appropriate frame and shifting between frames are determining factors in students' success at problem solving \cite{Scherr2009, Irving2013framing, Bing2012}.  Productive problem solving requires both an appropriate frame \cite{Scherr2009} and appropriate shifting between frames\cite{Nguyen2016StuFraming}.  Careful observation of student behaviors, gaze, and discourse can provide clues for determining students' epistemological frames.

In contrast to difficulties research, which seeks to find the wrong answers which prevent students from being successful at problem solving, framing research tends to focus on the pathways that students take as they shift frames to solve problems.  This emphasis on pathways rather than stopping points nudges framing research to more overtly consider the effects of problem statements, instructor interventions, and groupmates discourse on students' problem solving.

In our prior work, we developed a set of four inter-related frames around the idea of math-in-physics (Figure \ref{fig:framework})\cite{Modir2017QuantumFraming,Nguyen2016StuFraming}. We applied it to observational data to model students' framing in math and physics during in-class problem solving in two upper-division courses: quantum mechanics\cite{Modir2017QuantumFraming} and electromagnetic fields (E\&M)\cite{Nguyen2016StuFraming, ChariInsFraming}.  Briefly, our math-in-physics frames capture students' framing in math and physics, expanded through the algorithmic and conceptual space of students' problem solving. The four frames are: algorithmic math, conceptual math, algorithmic physics, and conceptual physics. We briefly characterize each frame as follows: 

\begin{figure}
\includegraphics[width=\figwide]{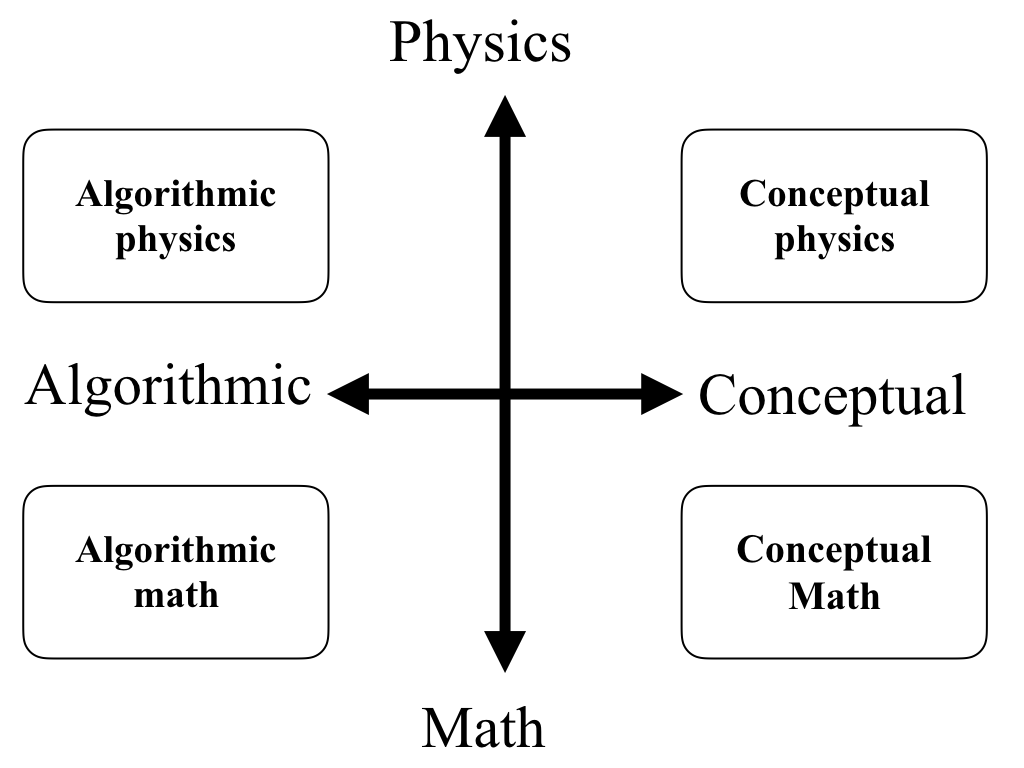}
\caption{Math-physics-algorithmic-conceptual theoretical
framework. The horizontal axis indicates the algorithmic and
conceptual directions. The vertical axis represents the math
versus physics directions. Each quadrant is labeled. Figure originally from \cite{Modir2017QuantumFraming}. \label{fig:framework}} 
\end{figure}

\begin{description}
\item[Algorithmic mathematics frame] Students are in an algorithmic mathematics frame if they think about mathematics algorithmically, e.g. when students do pure mathematical manipulations, such as taking a derivative, or checking for sign errors in their procedural problem solving. One of the hallmarks of algorithmic problem solving is that it is fast. Students in this frame take several fast and trivial  steps over a long period of time.

\item[Algorithmic physics frame] Students are in an algorithmic physics frame if they think about physics algorithmically, e.g. when laying out physics definitions by using mathematical formalisms. Additionally, students might only use an algorithmic heuristic to find a physical relation without writing down mathematics and only stating their reasoning verbally. This further clarifies the difference between algorithmic mathematics and algorithmic physics frames. 
 
\item[Conceptual mathematics frame] Students are in a conceptual mathematics frame when they provide reasoning, based on the properties of mathematical functions. Instead of running through the math algorithmically, students reason based on the general class of information in mathematics, e.g. when students notice an integral is equal to zero, without explicit calculations, and only due to identification of the mathematical properties of the integrand.

\item[Conceptual physics frame] Students are in a conceptual physics frame when they think about the features of a physical system, or think conceptually about physical laws, or explain a concept. Students may use graphical representations to better visualize the physical system. By taking a conceptual approach, students create more sense-making opportunities with less need for writing several algebra-based steps.

 \end{description}
Using this set of frames, we looked for moments where students' problem solving is impeded because they are in an unproductive frame or when a problem statement requires shifting between frames and students are unable to make that transition.
%BM-060119 removed in response to comment #4 of the reviewer:
%This paper applies the math-in-physics frames to secondary analysis of quantum mechanics difficulties. 
We present a mapping of over thirty quantum mechanics difficulties from the literature\cite{Singh2015, Emigh2015, Passante2015, Singh2008} to our math-in-physics frames.  This secondary analysis provides a deeper underlying structure to the reported difficulties and demonstrates the broad applicability of these frames to many kinds of quantum mechanics student data. 
%BM-060119: removed in response to reviewer's comment #4:

%However, our work in this paper is limited to the amount of information that is provided in the difficulty papers. We have excluded examples that we do not have enough evidence to map the difficulty into our categorizations. 

These categories represent a rotation of the basis set for student difficulties.  We have remapped the space of difficulties, which is loosely grouped by physics topics, into the space of frames, seeking an underlying cognitive structure.

% In section \ref{NEI}, we have defined a category called ``not enough information'' for those examples that we do not have enough evidence to map the difficulty into our categorizations.  These categories represent a rotation of the basis set for student difficulties away from specific topics in quantum mechanics and towards underlying cognitive structure.

\subsection{Interactions between problem statements and responses}

Students' reasoning comes as responses to specific questions, and those questions strongly influence their framing.  We examined problem statements for what frame(s) they initially promote. For example, consider these two problems:
\begin{itemize}
\item Using the time-independent Schr{\"o}dinger equation (TISE), calculate the changes to $E_{0}$, the ground state, as a given well shrinks from $L$ wide to $L/2$ wide.
\item What happens to the energy of the ground state when a finite square well gets narrower?
\end{itemize}
The first problem encourages students to think mathematically (``calculate'') and algorithmically (by hinting at a procedure).  The second is more conceptual, specifying neither numbers nor procedures.   These two problems, because of their different phrasing, may elicit different student difficulties, or the difficulties they elicit may appear in different proportions because of their phrasing.  Depending on which difficulties the researcher focuses on, the researcher might prefer one problem statement over another.

A major element of difficulties research is to carefully craft problem statements so as to best elicit student difficulties.  To honor this careful work, our secondary analysis of difficulties considers difficulties as they are paired to problem statements.  It's unusual for reports of difficulties research to closely examine how the interaction between question posed and students' ideas creates the student response (even as actual research work to elicit difficulties requires careful attention to question craft, the reports do not focus on it), so our insistence on examining questions paired with responses may seem strange to some difficulties researchers.  However, it is required by the framing framework, which sees students responses as fundamentally context-dependent.

\section{Methodology}

\subsection{Difficulty identification \label{sec:difficultiesmethods}}
Because this paper reinterprets existing datasets using new theory, we first review where the data come from and how they were originally analyzed using a difficulties framework. 

Researchers in difficulty studies have multiple methods for data collection, both quantitative and qualitative. The populations of students in these studies are drawn from advanced undergraduate courses and first-year graduate courses at several different US universities. Students are administered a written test (as part of their course work or for research purposes), usually at the beginning of the semester\cite{Singh2015} or after relevant instruction\cite{Passante2015, Emigh2015}. Some students also participate in think-aloud interviews intended to both develop the test and discover common responses to it. Data analysis on the interviews and written responses extracts common difficulties despite the differences in the students' backgrounds. The results of the analysis from interviews and tests are consistent. Several cycles of test development and administration adjust the questions to best elicit student difficulties and ensure validity and reliability.

The first paper, covered a broad range of data: administered surveys to upper-level undergraduate and graduate students simultaneously at several university across US, including 100-200 students; administered surveys to students in a typical  upper-level class size at state universities; and think-aloud interviews with students for in depth analysis\cite{Singh2015}.

The original data in two papers\cite{Passante2015, Emigh2015} were collected at the University of Washington (UW), where undergraduate physics students are required to take between one and three quantum mechanics courses. The first course (sophomore-level) covers the first five chapters of McIntyre's textbook\cite{Mcintyre2012QM}; the second and third courses (junior-level) cover all of Griffiths' textbook\cite{Griffiths2005QM}. Students were given a written pretest before relevant tutorial instruction, but after lecture instruction. In some of the tasks a variation of the questions was given to the students, but those question variations were not published.  

In the fourth paper\cite{Singh2008}, survey data of first-year graduate students were collected from seven different universities. Researchers also conducted interviews with fifteen students at the University of Pittsburgh. 

The research groups at both the University of Pittsburgh and the University of Washington have long histories of difficulties research in quantum mechanics and other physics subjects, and their expertise in developing questions, developing tests and curricula, and identifying difficulties is second to none. We chose their papers for secondary analysis because they represent the best that difficulties research in quantum mechanics has to offer.
 
\subsection{Mapping students' difficulties to framing\label{sec:secondary}}

We posit that many student difficulties in quantum mechanics may be due to unproductive framing in problem solving, because students' current frame may not help them with actual problem solving, because students find themselves temporarily unable to shift to a more productive frame, or because they cannot activate productive resources within their current frame. 
%To investigate this postulate, we conducted a secondary analysis of published student difficulties in quantum mechanics.   

We mapped descriptions of published difficulties into errors in epistemological framing and resource use. 

% BM-060119: removed: 
%Framing is context dependent and problem statement is one of the many contexts available to students in a problem solving situation. 

% BM-052519: I added some sentence in response to reviewer comment #2 on the role of problem statement as an important context

% BM-060119: removed: 
%From an epistemological view, student thought processes can be affected by different external and internal factors. 
Framing is context dependent, and a problem statement is one of the very first contexts students interact with which can influence students thought processes and even their future decision-making during a problem-solving scenario. 
A student's frame can be affected by different external frames. For example, in an individual problem-solving setting, the problem statement is one of the influential contexts. By contrast, if we consider a group problem-solving setting, the context can possibly expand from the problem statement to other students’ frames in the group. Furthermore, if the problem solving occurs in an interactive classroom setting, where the instructor occasionally intervenes to give a hint or further explain concepts to the groups, then the instructor frame can also affect and interact with students’ frames. In this study, the external factor that we are focused on is the problem statement, as the difficulties are elicited mostly in (1) individual problem-solving settings via think-aloud interviews with little or no intervention on the part of the interviewer, or in (2) written-mode settings such as surveys.

%Framing is what makes resources activation coherent and framing depends on the context. Students potentially, have a myriad of resources available to them to activate in a situation, however they do not activate their resources in a random manner. The mechanism that explains how students decide to which subset of resource or toolbox to access and activate in coherency is framing. 

%The external factor that we are focused on at a methodological level is the problem statement.  
 We considered the problem statement as the ``jumping off'' point for student framing, reasoning that students initial problem framing is probably strongly influenced by framing in the problem statement. From published descriptions of student responses -- including their written responses, where available -- we identified students response frames and compared them to the frame of the problems to categorize errors. 

Because this is a secondary analysis, we take the difficulty as the unit of analysis, not an individual student's response. This is a practical choice on our part, as some authors do not identify the frequency of each difficulty, and we did not have access to all of the descriptions of students' problem solving. The numbers reported for the error rates indicate how many difficulties fall in each category and do not indicate how many students have difficulties in each category.  This kind of analysis is strange in the knowledge in pieces research tradition, as it severely hampers us from looking at what students do that is correct or productive; difficulties-focused research does not report productive ideas, only incorrect ones.
% * <jeremy79@ksu.edu> 2018-11-26T19:04:10.256Z:
% 
% > This kind of analysis
% The kind of analysis where you indicate how many difficulties fall in each category? the kind of analysis where you indicate how many students have difficulties under a category? Both? Something else? *This* is unclear here.
% 
% ^.
% * <jeremy79@ksu.edu> 2018-11-26T19:02:16.379Z:
% 
% >  This is a practical choice on our part, as some authors do not identify the frequency of each difficulty
% I made an edit to get rid of an apparent redundancy, but I am also not sure about the meaning here.
% 
% 
% ^.

\subsection{Methodology for secondary analysis}

\subsubsection{Selection criteria}

We gathered published works which describe student difficulties in quantum mechanics from \textit{Physical Review Special Topics -- Physics Education Research, Physical Review -- Physics Education Research}, and the \textit{American Journal of Physics}.  We identified four papers and thirty-six student difficulties in quantum mechanics.

From these papers, we sought difficulties in which the authors had sufficiently described their problem statements (or instructor interactions) for us to determine initial problem framing, excluding those difficulties whose problem statements were omitted, or where variations on a problem statement were alluded to but not presented.  

There were times when our research team came to a consensus that there was not enough information to determine difficulties' probable framings. (Difficulties on problems for which the problem statement was not reported in enough detail are excluded from our analysis altogether). Out of thirty-six difficulties, twenty-seven difficulties remained. Nine difficulties did not have enough information for us to figure out what the framing could have been. We excluded these difficulties from further analysis, simply because there was not enough context to determine students' reasoning frames. 
% * <jeremy79@ksu.edu> 2018-11-26T19:23:04.221Z:
% 
% I moved this paragraph from Limitations. This eliminates some redundancy, and clarifies why you reduced from thirty-six to twenty-seven difficulties within the Selection criteria section.
% 
% ^.

%For our secondary analysis we had limited access to all environmental aspects of the literature's data sets. In some of these works the presence of the TA could have affected student's framing; however, there is not enough information to include the role of the TA in the secondary analysis. We will show that re-analysis of students' narrations that were being identified under the same difficulty topic can result in different ways of framing of the situation. This gives a finer grain analysis of students' cognitive process. In some of the problems when the variation of the problem was given to the students, the authors had not mentioned the problem statement, thus we did not include those descriptions in our analysis. 
%Due to the inherent limitations of our secondary analysis we proposed a method based on the presented information in the literatures to categorize and seek an underlying structure to the long lists of difficulties that span many topics in quantum mechanics.

\subsubsection{Coding}

We examined students responses with respect to the features of four frames in our math-in-physics set and coded the student framing present. We started with student responses to problems that matched our qualitative data\cite{Modir2017QuantumFraming}.  Descriptions of student responses -- and the resulting difficulties identified by researchers -- matched our observational qualitative data well. The detailed analysis of examples from our own observational data is not within the scope of this study, but can be found in section VI.A of the preceding published study\cite{Modir2017QuantumFraming}. In that study, using the lens of our developed theoretical framework, we identified students' frames and transitions in frames from analyzing in-class group problem-solving activities, as video-recorded in a senior-level quantum mechanics class\cite{Modir2017QuantumFraming}. 

Emboldened, we extended our coding of student responses to difficulties not present in our qualitative data. As much as possible, we investigated students' statements (or equations, on occasion) to identify the nature of their reasoning and frames used in students responses. For example, a response which is just a piece of an equation,  or an equation that is used as a plug-n-chug tool suggests that the student used an algorithmic frame to generate their response, whereas a response which coordinates energy and probability descriptions suggests that the student argued from a physical principle and is in a conceptual physics frame.

%Our framework was comprehensive enough to conduct a secondary analysis on the students' responses reported in the literature. Surprisingly, we found instances in the responses of the students and our own data that framed a similar situation accordingly. Additionally, we found instances of students' activities that have been not seen in our data; however, we were able to successfully find evidence to map them to our framework. This expanded our coding scheme in characterizing each of the quadrants based on a broader range of the same class of activities. 
%
%We investigated the statement of the students to identify the nature of their discussion, or analyzed the artifacts in their responses when they had been just focusing on an equation, and no further narration was given by the student. For example a response which is just a piece of an equation, suggests the features of an algorithmic frame where there is less argument compared to other quadrants. 
 
%\subsection{Problem statement coding}
To investigate how problem frame affects student frame, we  coded for frames promoted
by problem statements. We looked at the keywords and the givens in the problem statement to identify the starting frame(s) suggested by the problem statement. 
Some problem statements, particularly multi-part problems, require students to start in one frame and shift to another one (for example, see section \ref{sec:Ex1}).  In those cases, we coded for which sequence(s) of frames would yield correct answers.

%\subsubsection{Reliability testing}

Through intensive discussion among multiple researchers,  we coded for which frame(s) a problem statement promoted, and which frame(s) were evident in students' reasoning.  For some difficulties, responses or the descriptions of students' reasoning did not contain enough detail to figure out students' framing.  Our goals in these discussions were to come to agreement about our inferences of student reasoning. As our discussion reached consensus and our codebook stabilized, two independent raters coded both the rating of the problem statements and the ratings of the students' responses (or the descriptions of students' reasoning), with an agreement rate of $>90\%$ for both kinds of coding.

\subsubsection{Error type determination}

\begin{figure}
\includegraphics[width=\figwide]{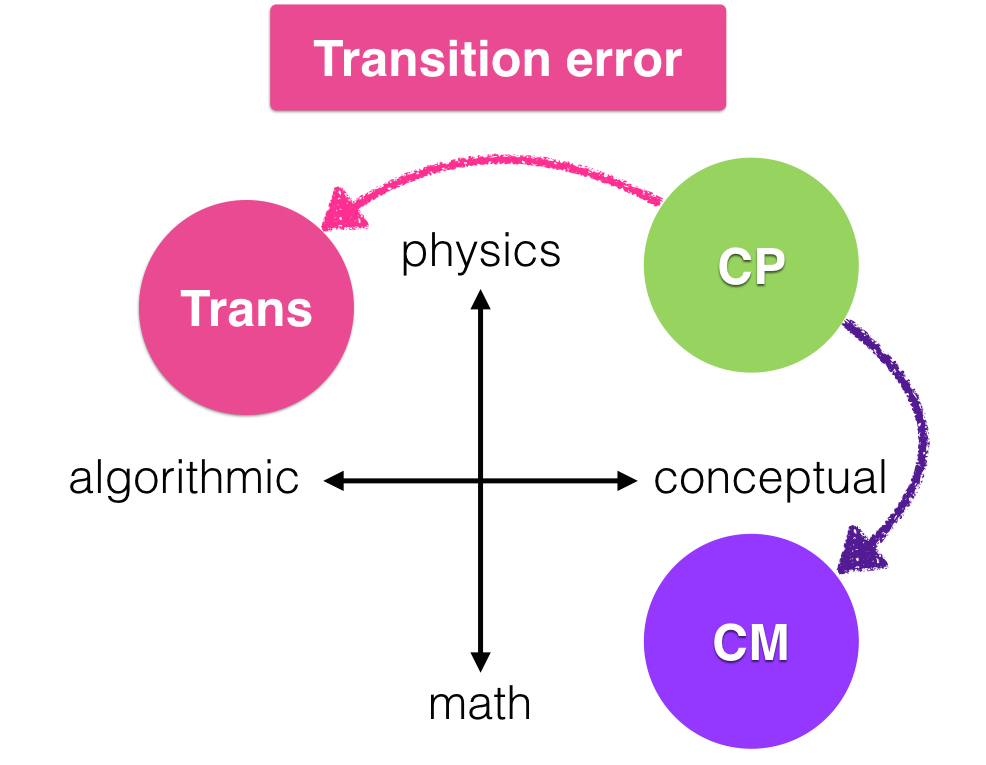}
\caption{Transition error. CP stands for conceptual physics frame, CM stands for conceptual mathematics frame, Trans stands for transition error. This student starts in CP and should transition to CM, but has instead transitioned to algorithmic physics.} \label{fig:Trans} 
\end{figure}

\begin{figure}
 \includegraphics[width=\figwide]{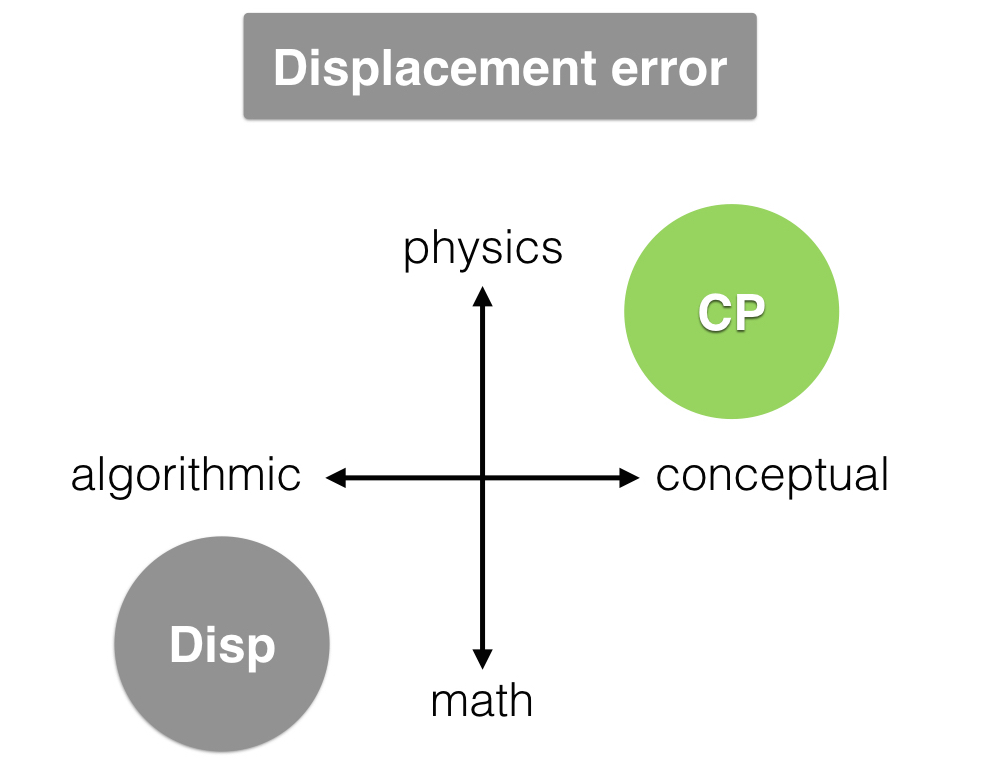}
 \caption{Displacement error. Disp stands for displacement error.  This student should be in in CP, but is in algorithmic math.}  \label{fig:Disp}
 \end{figure}

\begin{figure}
\includegraphics[width=\figwide]{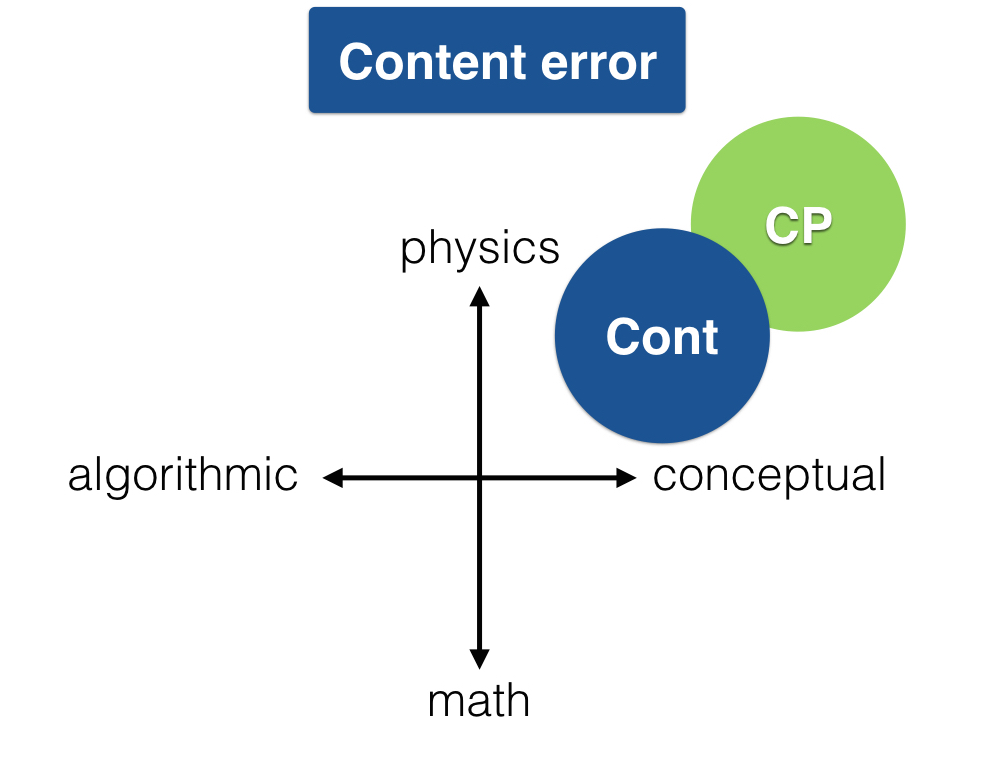}
\caption{Content error. Cont stands for content error. The student is in CP appropriately, but is using inappropriate or insufficient resources. } \label{fig:Con}
\end{figure}

Once problem statements were coded for frames promoted and student responses were coded for frames used, we classified students' difficulties into three categories:
\begin{description}
\item[Transition error] when a problem statement requires shifting from one frame to another, and students are unable to make a productive transition. (Figure \ref{fig:Trans})
% * <jeremy79@ksu.edu> 2018-11-28T13:10:26.033Z:
% 
% > when a problem statement requires shifting between frames, and students are unable to make that transition.
% As previously written, this implies that transition error is a matter of getting stuck in a particular frame, but Figure 2 and other description makes it clear that transition error can also be a result of shifting to an unproductive frame. The changes made are intended to help clarify this point.  
% 
% ^.
\item[Displacement error] when a problem statement promotes one frame, but students' reasoning places them in another frame. (Figure \ref{fig:Disp})
\item[Content error] when students appear to be framing the problem correctly, but are not activating appropriate resources to solve it. (Figure \ref{fig:Con})
\end{description}

This naming scheme relies heavily on the metaphor of framing as a location in a plane. In other words, transition error is {\it going to} the wrong place, displacement error is {\it being in} the wrong place, and content error is {\it being in the right place but using} the wrong ideas. 
We have illustrated each error visually in Figures \ref{fig:Trans}-\ref{fig:Con}, by considering a hypothetical problem statement that promotes the conceptual physics frame as the starting frame, and requires transition to the conceptual mathematics frame.

Figure \ref{fig:Trans} demonstrates a transition error for a hypothetical student that does not go to the right place (conceptual mathematics frame). 
%activate resources in one frame, but cannot shift to another frame, or transition to the wrong frame.
Figure \ref{fig:Disp} depicts a displacement error when the student is in the wrong place (algorithmic mathematics frame). 
% are not in the frame intended by the problem statement. 
Figure \ref{fig:Con} shows a content error when a student is in the right place (conceptual physics frame), but using wrong ideas. 
%are in the frame intended by the problem statement, but do not activate appropriate resources. 

\subsubsection{Limitations}
Many student responses to these questions are correct, and our secondary analysis of student difficulties cannot capture those responses.  This is a fundamental limitation of difficulties-based research: it seeks to describe the ways students are wrong, not the ways that their responses are reasonable.

Some difficulties may have arisen due to multiple types of error.  This is a limitation of secondary analysis -- we do not have full reports of student reasoning -- and of the survey-style free-response data on which many of the original difficulties are based.  For this reason, we classified some difficulties as arising from multiple error types.  With sufficiently detailed data, we believe that each difficulty-displaying student response can be classified into a single error type.  

Additionally, some surveys were multiple choice.  While the original researchers based the choices on common student reasoning, and that reasoning could have showed evidence of student framing, the multiple-choice answers themselves are often insufficiently detailed to determine students' framing.  As much as possible, we coded researchers' descriptions of student difficulties, but sometimes we simply did not have enough information.   

In the following sections, we show examples of each kind of error, arguing 
%from our data and 
from published literature that difficulties can be categorized by framing error type. Within each type, we tabulate published difficulties. Because some problems require transitions between frames and some do not, we classify difficulties first by the kind of problem they come from and second by the kinds of errors they produce. More specifically, for each framing error type, the error categorization is first provided for problems that require transitions, followed by simpler problems that do not require transitions. The table arrangements also appear in the same order. Error types are labeled by capital letters; the small letters stand for the kind of problem.

\section{Transition error}
Transition errors occur when a problem statement expects students to shift between two frames, and the students either do not shift, or shift into an unproductive frame.  In this section, we first motivate the idea of transition errors through extended analysis of one example, then tabulate all difficulties which exhibit transition errors.

\subsection{Transition error example\label{sec:Ex1}}
\begin{figure}
 \includegraphics[width=\figwide]{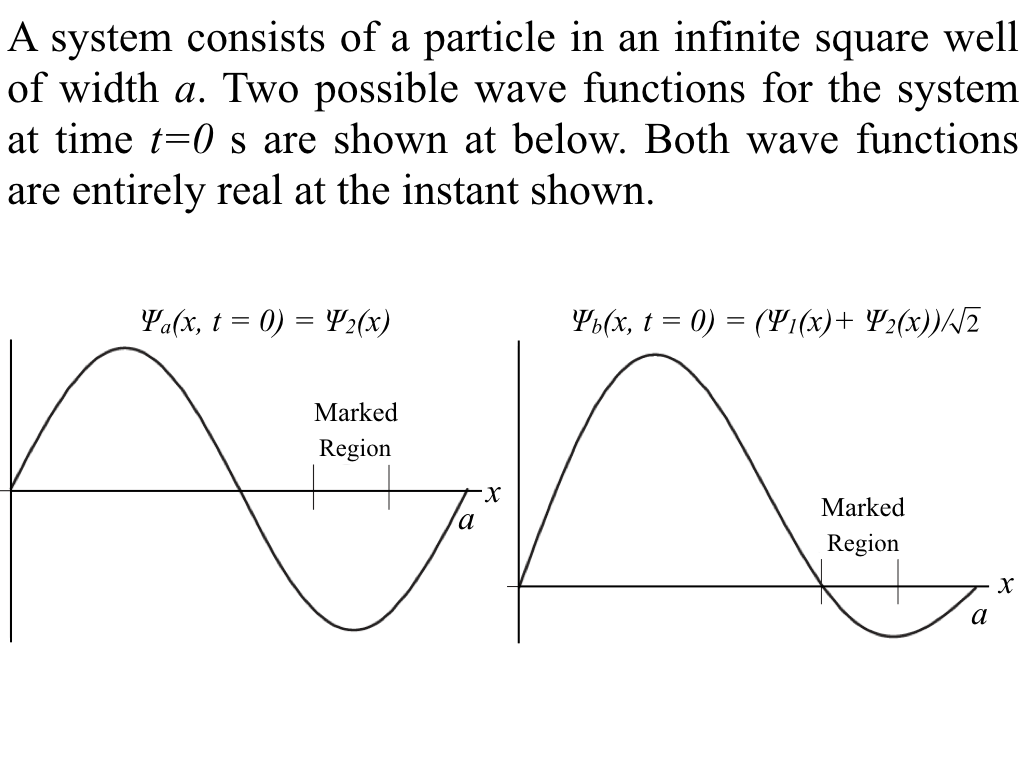}
 \caption{Does the probability of finding a particle in the marked region depend on time? In this problem, student difficulties indicate  transition error and displacement error. Figure originally from \cite{Emigh2015}. \label{fig:Emighgraph}} 
 \end{figure}

The first example illustrates a transition error which arises from interpreting a graph of wave function vs position\cite{Emigh2015} (Figure \ref{fig:Emighgraph}). The problem asked students to explain if the probability of finding the particles within a marked region depends on time or not.

The probability density depends on time if the modulus square of the wave function depends on time. The wave functions are given at time $t=0$. The authors mention that the problem requires students to think about the time dependent phase of each term in the superposition wave function. This encourages students to frame the question as thinking about what it means to be in a superposition of states, what are the energies of each term in the superposition, and how does the system evolve over time.  
%Additionally, the problem statement includes multiple representations for students to conceptually coordinate. 
Framing the problem in this manner suggests conceptual physics as the initial frame.

Students may start this problem by thinking of the time evolution operator, which is determined by the Hamiltonian of the system.
After students recognize the correct time-phase factors, they need to coordinate mathematical representations to show how the phase factors determine the time dependence of the probability density. 
In the context of quantum mechanics with more novel mathematical formalism, students can use mathematics in epistemologically different ways to map their physical understanding to a mathematical representation. 
% * <jeremy79@ksu.edu> 2018-11-26T19:36:50.885Z:
% 
% > in an epistemologically different manner 
% edit made to eliminate double-adverb, which wasn't scanning well
% 
% ^.

In algorithmic problem solving, the mathematical process is broken into many smaller algebraic steps and taken over a longer period of time. Whereas, in conceptual problem solving, a mathematical justification can account for all, or part of the algorithmic steps. Sometimes, in a problem, students need to make a transition between algorithmic and conceptual mathematics to fully coordinate all the features of a physical system into a mathematical representation. 
Our set of math-in-physics frames has two possible transitions from conceptual physics into mathematics frames: 
\begin{description}
\item[Algorithmic math] In algorithmic math, a student would manipulate the modulus square of the superposed wave function explicitly and algorithmically, finding that the time dependence of the pure terms falls out, and the time dependence of the cross terms persists.  
\item[Conceptual math] In conceptual math, a student initially could use a conceptual mathematical shortcut: the exponential term multiplied by its complex conjugate sets the product equal to one.  %This was Eric's initial framing in Section \ref{sec:Eric} on a comparable problem.  
However, %just as Eric noticed, 
this solution leads to neglecting the role of the cross terms.
\end{description}
Because the problem starts in a conceptual physics frame, it may be easier or more appealing to transition first into conceptual math rather than algorithmic math; in our observational data, this is the transition we observed\cite{Modir2017QuantumFraming} from
the student (Eric).  %Eric made that transition originally. 
Conceptual mathematics was Eric's initial framing on a comparable problem.  Using conceptual mathematics reasoning just as Eric did can only explain that the pure terms lose their time dependence. Eric then made a transition to algorithmic math,  allowing him to mathematically read out that in computing the modulus square of the probability density there are cross terms and for those terms the time dependence persists. 

Emigh et al. \cite{Emigh2015} describe student reasoning in response to the same task: 
 \begin{description}\label{studEmighstudy}
\item[Student] While it is true that the general wave function is of the form $\sqrt\frac{1}{2}\phi_{1}e^\frac{-iE_{1}t}{\hbar}$  $+$ $\sqrt\frac{1}{2}\phi_{2}e^\frac{-iE_{2}t}{\hbar}$ again the function we are interested in is $ P(x) = \mid \phi \mid^2$ which loses its time dependence.  
\end{description}

The first part of this statement shows that the student has correctly used the ideas of the problem statement frame to note the different energies of each energy eigenstate in the wave function. The second part of the statement suggests that the student coordinates the physics and conceptual math to recall that the probability density is the modulus square of the wave function. However, the student does not do any further algorithmic calculations, instead arguing that the probability ``loses the time dependence''. This is congruent with the conceptual math reasoning above. Eric's initial frame is similar to the student in the Emigh et al study in that he only accounts for part of the solution. This leads to an error without Eric being aware of it.   
%BM-052819:Added in response to reviewer comment #3:
According to both the difficulties framework and knowledge in pieces, at this moment an error has occurred in the problem-solving procedure. However, by using a framing lens we can extend our analysis further to a level that incorporates the role of other existing external factors in a real problem-solving situation such as a group problem-solving classroom setting. 
It is only after the instructor gives the correct final answer to the class that Eric becomes aware of his error, which had momentarily made him `get stuck'. Eric is able to `get unstuck' by making a transition to the algorithmic frame and paying attention to the features of the problem that are easy to see in the new frame.

%and also with Eric's initial reasoning in Section \ref{sec:Eric}.
 
About $10-20 \%$ of the students (N=416) from Emigh et al., applied the same kind of reasoning to argue that ``time drops out'' or ``the probability is squared and the time won't matter'' \cite{Emigh2015}. These arguments indicate that students do not feel a need to actually do math, because their conceptual math frame has solved and justified their time-dependent answer. While Emigh et al interpreted these responses as a difficulty -- students' ``tendency to treat all wave functions as having a single phase'' -- we interpret it as an example of error in frame transition (Table \ref{tab:5}).  

\subsection{Difficulties which exhibit transition errors}

We found two published difficulties for which students exhibit only transition errors (Table \ref{tab:3}).

\begin{dtable}{Difficulties that exhibit transition error only.\label{tab:3} These difficulties are labeled ``T'' for transition errors. The ordering of the difficulties in the table is for labeling purposes only and does not represent a hierarchy.}
T1 & Mathematical representations of non-stationary state wave function; this difficulty emerges when determining possible wave functions for a system. & \cite{Singh2015}\\
\hline
T2 & Belief that the wave function is time independent; this difficulty emerges when interpreting the time phases that arise from the time dependent Schr{\"o}dinger equation.  & \cite{Emigh2015}\\
\end{dtable}

In the first difficulty \textit{mathematical representations of non stationary state wave functions} (T1) in Table \ref{tab:3}, the students were asked if different wave functions: $A\sin^3({\pi}x/a)$, $A[\sqrt\frac{2}{5}\sin({\pi}x/a)+\sqrt\frac{3}{5}\sin(2{\pi}x/a)]$ and $Ae^{-(\frac{(x-\frac{a}{2})}{a})^2}$ can be proper candidates for an infinite square well of width \textit{a} with boundaries at $x=0$ and $x=a$. This problem requires students to start from a conceptual physics frame to extract the boundary condition information and read out that the potential is infinite at the boundary conditions and the wave function has to go to zero to satisfy the continuity of the wave function at the boundaries. This problem requires transition as students may need to plug in the values for the boundary conditions, conduct some algorithmic steps, and figure out if the solution satisfies the boundaries of the problem. 
% * <jeremy79@ksu.edu> 2018-11-26T19:48:10.393Z:
% 
% I deleted the statement here, because it didn't seem to add anything. That said, if you decide to leave it in, I recommend that you clean up the wording. As is, it's very awkward. "Students often exhibit transition errors when attempting this problem." would be fine and say the same thing. 
% 
% ^.
One type of incorrect response suggests that many of the students reasoned that two conditions must be satisfied. First, wave functions should be smooth, single valued, and satisfy the boundary conditions of the physical system.  Second, it should be possible to write the wave function as a superposition of stationary states, or the wave function should satisfy the time independent Schr{\"o}dinger equation (TISE)\cite{Singh2008}. 
A typical response of the students looks like: 

\begin{description}
\item[Student]  $A\sin^3({\pi}x/a)$ satisfies b.c. but does not satisfy Schr{\"o}dinger equation that is, it cannot represent a particle wave. The second one is a solution to S.E. it is a particle wave. The third does not satisfy b.c.
\end{description}
The author mentions that students do not note that even the superposition wave function $(A[\sqrt\frac{2}{5}\sin({\pi}x/a)+\sqrt\frac{3}{5}\sin(2{\pi}x/a)])$ does not satisfy the TISE.
We think that this student is in the frame of the problem since they match the boundary conditions with each wave function to see if they satisfy the boundaries of the physical system. However, we do not have sufficient information to conclude how this student is working on this problem towards a solution. We do not know if the student is taking some algebraic steps to match the boundary condition, or if they are only reasoning verbally.
% * <jeremy79@ksu.edu> 2018-11-26T20:01:17.028Z:
% 
% > However, we do not have sufficient information to conclude that is how this student is working on this problem as part of their solution. We do not know if they take some algebraic steps to match the boundary condition, or if they only reason verbally?
% In attempting to clarify this statement, I might have changed its intended meaning. (Please check!) Regardless, the original statement was awkward to comprehend, and needed clarification. 
% 
% 
% ^.
% ^.
We think that, by making a transition to the conceptual mathematics frame, this student can activate ideas regarding expansion of the function $\sin^3(\pi x/a)$ in terms of the energy eigenstates.%\ecs{check this claim.  Is this what we mean?}

A second difficulty is that students believe the wave function is time independent because it satisfies the TISE. Students who generate these responses provide a mathematical basis for their answer. The author mentions that students think that the superposition wave function satisfies the TISE. 

 \begin{description}
\item[Student] [both wave functions] satisfy the time independent Schr{\"o}dinger equation so $\Psi_1$ and $\Psi_2$ do not have time dependence. 
\end{description}
Although the solution to the TISE does not depend on time, the TISE solution is incomplete because this problem is time-dependent. This difficulty is categorized as a transition error, as students need to shift to a conceptual frame (either conceptual physics or conceptual mathematics) to complete the problem. Shifting to conceptual physics may lead them to think in terms of the independent eigenfunctions of space and time; shifting to conceptual math may lead them to think about missing orthogonal functions.

This problem may also exhibit displacement errors. To better allow the reader to compare displacement error with the transition errors described above, we present our analysis of two possible displacement errors emerging from this problem statement in the follow section (subsection \ref{sec:disptrans}). 

\section{Displacement error\label{sec:Ex2}}
Displacement errors arise when students are meant to be in one frame, but instead operate in another.  In this section, we first describe a displacement error that a student may exhibit when attempting the extended example problem in the prior section.  Then we tabulate difficulties which exhibit displacement errors for problems with and without transitions.
% * <jeremy79@ksu.edu> 2018-11-27T14:19:20.182Z:
% 
% "as if the student exhibited a displacement error" is confusing. I'm not sure if it means you have the same observations of the student, but interpret those observations as being the result of a displacement error instead of a transition error, or if you have different observations that result in you concluding that the student is making a displacement error, and not a transition error. I rewrote this sentence to clarify that you mean the latter (observing different student behavior) based on the later description. 
% 
% ^.

\subsection{Displacement error example\label{sec:disptrans}}
For the same task as shown in Figure \ref{fig:Emighgraph} (the probability of finding a particle in the marked region), we present two possible displacement errors.

Some students considered that just the linear combination $A[\sqrt\frac{2}{5}\sin({\pi}x/a)+\sqrt\frac{3}{5}\sin(2{\pi}x/a)]$ or a pure sinusoidal wave function are allowed; but the $A\sin^3({\pi}x/a)$ is not allowed and ``only simple sines or cosines are allowed" as proper wave functions. Some other students mentioned that for a particle in a box, only the wave functions in the form of $A\sin({n{\pi}x/a})$ are allowed and the $Ae^{-(\frac{(x-\frac{a}{2})}{a})^2}$ wave function is only allowed for a simple harmonic oscillator.

We consider that students with this type of response are not in the frame of the problem as they are not thinking about the characteristics of the boundary conditions. Instead, they just recall what the solutions for physical systems of a particle in a box or a harmonic oscillator look like. They assert that they know how the answer should look, having worked out the problem before. While the students might not necessarily attempt the algorithmic processes to arrive at this conclusion during the interview, they are relying on the fact that they have done these calculations before and can recall the conclusion. This tendency stems from having previously worked through the problem of a particle in a box, which fits into the algorithmic physics frame. We categorized students with this type of response as having made a displacement error.

Other difficulties displacement errors are possible.  For example, this student writes the time dependence of the wave function instead of finding the probability, which is an incomplete answer:

 \begin{align}
\sqrt\frac{1}{2}e^{-iEt/\hbar}(\psi_{1}+\psi_{2}) 
\end{align}

This short answer segment suggests that the student is in an algorithmic frame; there is no other information about student reasoning (such as narration or a graph provided by the student). This student has not picked up the conceptual framing intended by the problem statement. Starting from the conceptual physics frame could help the student to conceptually think  about superposition of wave functions and the different energy terms instead of a single time dependent phase. The authors of the original paper\cite{Emigh2015} do not provide the percentage of students that answered in this way. However, they mention that the \textit{tendency to consider just a single phase wave function} for a superposition state is very common.  About 25\% of their students ($N=223$) on a final exam showed the same difficulty on a version of the same task. Students were given the time dependent wave function, and were asked  about the ``time dependence of the probability of a particular outcome of a position measurement".

This difficulty is classified as a displacement error: the student is in the wrong frame initially, and does not transition to a more productive frame. 

\subsection{Difficulties which exhibit displacement errors}
In problems which require frame shifting (like the problem in Figure \ref{fig:Emighgraph}), we found five difficulties which exhibit only displacement error (Table \ref{tab:1}). 

One difficulty from Emigh's study\cite{Emigh2015} (Problem in Figure \ref{fig:Emighgraph}), \textit{confusion between the time dependence of wave functions and probability density} (Ds2) in table \ref{tab:1}, shows that students correspond the time dependence of one quantity to another such that both physics quantities obtain the same time evolution. Between 5 and 20\% of the students in their data (N=416) have provided this type of reasoning.  %\mwscomment{this feels grammatically problematic.  Do you mean "One problem from Emigh's study" or "On problems from Emigh's study or something else?}

\begin{description}
\item[Student]  The wave function is time independent. Thus, its probability density does not change. If the wave function is time dependent, then [its] probability density would change in time too.

\end{description}
 
This student does not calculate the modulus square either via an algorithmic mathematical frame, a conceptual mathematical frame, or both. Additionally, the student does not think conceptually about the different energy eigenvalues of each term in the superposition. This student is not in the frame of the problem, which is a conceptual physics frame. Instead, the student is in the algorithmic physics frame. 
This student is using a simple algorithmic heuristic: If this thing (wave function) is not changing, the other thing (probability density) is also not changing; if this (wave function) were changing then its (probability density) would also be changing. The student is applying this algorithmic piece of reasoning to the physical quantities of (wave function) and (probability density), without considering the physics of those quantities deeply and conceptually. The student is applying algorithmic reasoning to make a quick conclusion about the relation between two physical quantities. One of the hallmarks of algorithmic thinking is that it is fast and non-reflective. 
% * <jeremy79@ksu.edu> 2018-11-27T14:37:02.691Z:
% 
% > and non-reflective
% I'm not quite sure if this is the correct term, but "fast" alone seems to be insufficient. It's not just that algorithms are fast, but that they allow the student to avoid reflecting on underlying concepts.  
% 
% ^.
It is possible to totally answer this problem conceptually and be wrong. But, we do not have further information such as student's tone or before and after this argument to investigate this possibility. 
%Without this context, our conclusions about the student's reasoning are limited. 

%:It is also possible that the student was in the conceptual physics frame when exhibiting the reasoning above. Unfortunately, we do not have further information about the student's tone or the rest of the arguments. Without this context, our conclusions about the student's reasoning are limited. 
% * <jeremy79@ksu.edu> 2018-11-27T14:41:45.223Z:
% 
% > It is possible to totally answer this conceptually and be wrong. 
% I did significant work on this paragraph, which might have changed the meaning. As it was, it was difficult to understand, with the first sentence of the paragraph not flowing well with the rest of the paragraph. Make sure that, whatever I did here, the intended meaning is still there. I think this paragraph is a general acknowledgement that, while the student was probably making a displacement error as they exhibited the reasoning described, there isn't enough context to be certain about that. As it was originally written, the first sentence of this paragraph seems to be about something else entirely: that it is possible to be in the conceptual physics frame, and still make errors on this problem. 
% 
% ^.
The piece of algorithmic physics reasoning given appears to indicate that the student is recalling, but it may not be true that all errors of this type arise from recollection. 
%This errors occur mostly when students just recall a fact about a physical quantity and they do not justify it. Usually in these cases, students do not interact with the problem statement to extract information out of it, because they think the recalled fact is unconditionally true in any context or physical system.

For the stationary state wave function on the same problem (Figure \ref{fig:Emighgraph}), about 5\% of the students think that the stationary state wave function is time independent.

\begin{description}
\item[Student]  This is a stationary state so the wave function will not evolve with time.

\end{description}

This piece of data suggests that this student is not initially in the conceptual frame of the problem. The student might have previously derived that some properties, such as probability density, are time independent for a stationary state. However, they do not accurately remember the conclusion. The incorrect notion that the wave function, rather than the probability density, is time independent further implies that the student is in an algorithmic physics frame and is trying to recall a fact about stationary states. 
 
%\BM{Comment 1: added}

In simpler problems that do not require frame shifting, we find one difficulty which exhibits displacement error (Table \ref{tab:9}).

\begin{dtable}{Difficulties which exhibit displacement error in problems that require frame shifting. \label{tab:1} These difficulties are labeled ``D'' for displacement errors and ``s'' because their problems require shifting. The ordering of the difficulties in the table is for labeling purposes only and does not represent a hierarchy.} 	
Ds1 & Incorrect belief that $H\psi=E\psi$ holds for any possible wave function $\psi$ & \cite{Singh2015}\\
\hline
Ds2 & Confusion between the time dependence of wave functions and probability density & \cite{Emigh2015}\\
\hline
 Ds3 & Belief that for a time-independent Hamiltonian, the wave function does not depend on time  & \cite{Singh2015}\\
\hline
Ds4 & Tendency to associate the time dependence of energy measurements with properties of stationary states & \cite{Passante2015}\\
\hline
Ds5 & Tendency to treat every superposition as having multiple distinct phases; this difficulty emerges when interpreting the time phases that arise from the time dependent Schr{\"o}dinger equation. & \cite{Emigh2015}\\
%\hline
%Ds6 & Tendency to treat superposition as having a multiple distinct phases & \cite{Emigh2015}\\
%\hline
%Ds7 & Difficulties in distinguishing between vectors in real space and Hilbert space & \cite{Singh2015}\\
%\hline		
\end{dtable}

\begin{dtable}{Difficulties that exhibit only displacement error in simpler problems which do not require frame shifting. \label{tab:9} These difficulties are labeled ``D'' for displacement errors and ``n'' because their problems require no shifting. The ordering of the difficulties in the table is for labeling purposes only and does not represent a hierarchy.} 	
%Dn1 & Incorrect belief that the time evolution of a wave function is always via an overall phase factor of the type $e^\frac{-iEt}{\hbar}$ & \cite{Singh2015}\\
%\hline
Dn1 & Difficulties related to outside knowledge in student understanding of energy measurements& \cite{Passante2015}\\
%\hline
\end{dtable}

% In simpler problems that do not require frame shifting, we find two difficulties which exhibit displacement error (Table \ref{tab:9}). These errors occur mostly when students are just in a remembering frame \ecs{remembering as a frame is new. Let's restrict our inquiry to only the 4 frames we've already got.} of mind to recall the properties of a physics quantity without justification, or further consideration of the problem statement. The task asks students to discuss the effect of perturbation on the time dependence of the probability density of a particle in a infinite square well. Emigh et al. showed that students confuse the time dependence of the potential of a perturbed system with the time dependence aspect of the probability density (Dn1)\cite{Emigh2015}. These kind of responses lack the relevant aspects of the quantum mechanics concepts. \ecs{why is this displacement error and not insufficient information?}

%\begin{dtable}{Difficulties that exhibit only displacement error in simpler problems which do not require frame shifting; and ``n'' because their problems require no shifting. \label{tab:9} These difficulties are labeled ``D'' for displacement errors; the ordering of the difficulties in the table is for labeling purposes only and does not represent a hierarchy.} 	
%Dn1 & Confusion between the time dependence of the potential and other quantities & \cite{Emigh2015}\\
%\hline
%Dn2 & Difficulties related to outside knowledge in energy measurements & \cite{Passante2015}\\
%\hline
%\end{dtable} 

 \section{Content error\label{sec:Ex3}}
A third kind of error occurs when students are in the appropriate frame intended by the problem statement, but have not activated enough of (or the correct) resources to complete the problem. We term this kind of error ``content error''.    In this section, we illustrate content error with one example difficulty and then tabulate content errors.
 
\subsection{Content error example}
\begin{figure}
 \begin{framed}
 \includegraphics[width=\figwide]{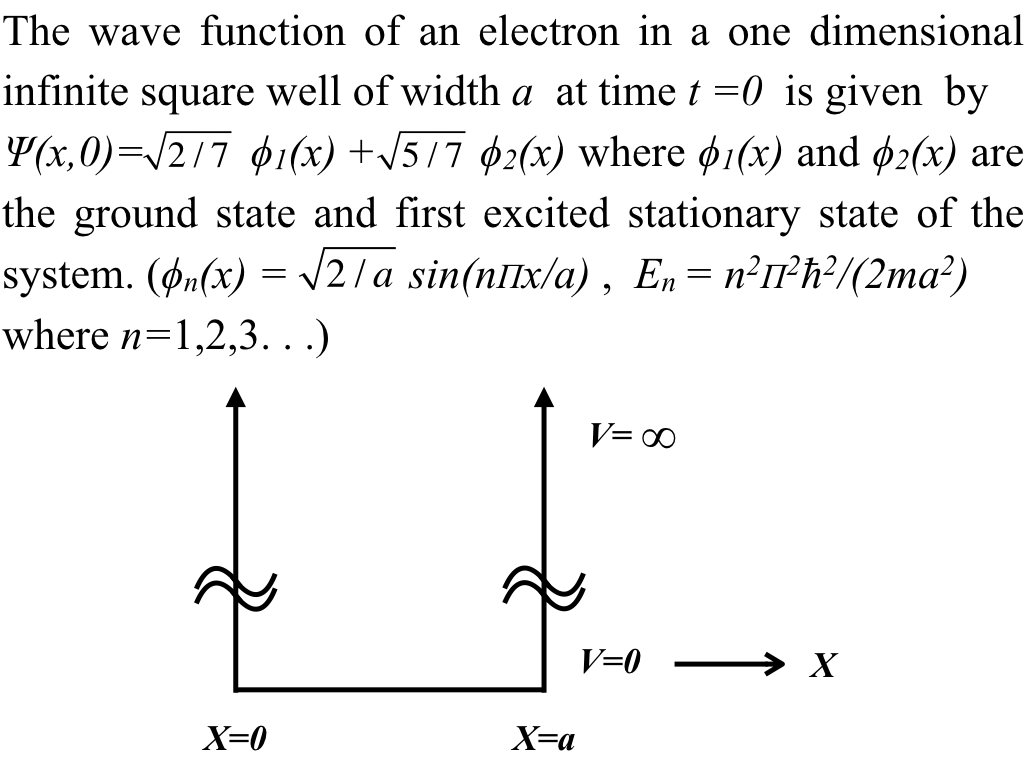}
\begin{enumerate}
\item Write down the time dependent wave function. 
\item You measure the energy of an electron at time $t = 0$. Write down the possible values of the energy and the probability of measuring each.
\item Calculate the expectation value of the energy in the state $(x, t)$ above.
\end{enumerate}
\end{framed}
 \caption{Calculate the expectation value of the energy in the state $\psi(x,t)$. On this problem, student difficulties display content error. Figure originally from \cite{Singh2008}. \label{fig:Singhgraph}} 
 \end{figure}

To illustrate content error, we draw an example from Singh's study (Figure \ref{fig:Singhgraph}) \cite{Singh2008}.  This example comes from the interview data of first-year graduate students. For this problem, students are given the problem in Figure \ref{fig:Singhgraph}, which asks them to calculate the expectation value of the superposition of the ground state and the first excited stationary state of the system. %Part (a) of the question asks to write down the time dependence wave function. And part (b) asks ``You measure the energy of an electron at time t = 0. Write down the possible values of the energy and the probability of measuring each''. And part (c) asks to ``Calculate the expectation value of the energy in the state x, t above.''

Although 67\% of the students were able to answer part (2) correctly, only 39\% were able to answer part (3) correctly, and many were not able to use the information to apply in part (3). Instead, students explicitly calculated the integrals of the expectation value.
% * <jeremy79@ksu.edu> 2018-11-27T15:15:42.218Z:
% 
% >  use the information
% The information given? The information gained from part 2? This is not clear.
% 
% ^.
We analyze their description of a student's response to part (3).  The problem statement starts students in an algorithmic frame (directing them to ``calculate'').  The frame is algorithmic physics, rather than algorithmic math, because the students must first start by recalling some facts and equations about expectation values and wave functions. 
% * <jeremy79@ksu.edu> 2018-11-27T15:18:34.984Z:
% 
% >  their 
% Who is "they" here? Singh et al.? It isn't really clear
% 
% ^.
%During the problem solving procedure student may require to make a transition to algorithmic math frame to conduct their calculations. 

The student writes down the TISE as $\hat{H}\phi_{n}=E_{n}$ without $\phi_{n}$ on the right-hand side of the equation, but correctly writes $\phi_n$ as the sum of $\phi_1$ and $\phi_2$ on the left-hand side. This is an appropriate initial framing to this problem, but it is missing a key piece of content. This mistake results in an incorrect answer in terms of $\phi_{1}$ and $\phi_{2}$. At this point, the student is not confused that their answer does not make sense because they are unaware of their error. The interviewer points to the part of the solution with the missing element, but the student is still unable to find their mistake. Finally, the interviewer explicitly gives the right TISE, $\hat{H}\phi_{n}=E_{n}\phi_{n}$ to the student. 
% * <jeremy79@ksu.edu> 2018-11-27T15:21:38.853Z:
% 
% > At this point, the student is not confused that his answer does not make sense because he is unaware 
% I switched to gender-neutral pronouns in throughout this section (for both the student and the instructor). The gender of the student and instructor seem to be immaterial to the description, and you have been using gender-neutral pronouns up to this point. Switching to gender neutral pronouns makes sense for consistency if nothing else. 
% 
% ^.
The student can then review the math conceptually in their solution by applying the orthonormality properties of the eigenstates, simplifying the integration, and getting a correct answer. It seems that all they need is a correct TISE, and they are able to frame the problem appropriately and continue to a successful solution. We do not consider this example as a case of a simple typographic error on the student's part, because the instructor notifies the student about their error several times, but the student believes that their written TISE is fine. After finding the correct answer, the student is able to reflect on their answer and even conceptually reason about the expectation value.

The interviewer continues by asking the student if they can think of the response to part (3) in terms of the response of part (2). The student responds ``Oh yes $\dots$I never thought of it this way$\dots$I can just multiply the probability of measuring a particular energy with that energy and add them up to get the expectation value because expectation value is the average value.''  The interviewer's intervention to explicitly connect parts (2) and (3) prompts the student to think more physically in terms of the underlying concept of expectation value.  They can relate the concept of expectation value to the parameter of the physical system such as energy eigenvalues, and probability of measuring each.
This difficulty is categorized as a content error because the student is in the frame intended by the problem but is not able to find the correct answer until the interviewer provides more content. In this example, the student in a brief but important interaction with the interviewer receives feedback and is able to continue from there to incorporate that given piece of information to revise the solution. 
Some students may also need help with the incorporation of the given information into their existing knowledge network; to add, remove, alter, or refine an idea(s) in their mind\cite{Hammer2003Tapping, Wittmann2006Conceptual}. 
This view can help instructors to better identify  the moments during a problem-solving situation to provide a piece of content to students or to help students to activate a piece of content and account for variation in students' reasoning. 

 %As discussed in the paper, many students get ``stuck'' by neglecting to take the complex conjugate and using orthonormality properties to simplify their manipulations. From a frame perspective the latter cases are different from the student in this example, as he is able to coordinate the representations and apply the meaning of the mathematical process at once; however, he is not able to proceed due to a missing factor in his solution. 
%The final conversation of the interviewer with the student indicates that the student finally is able to see the connection between the probability of energy values and finding the expectation value. This evidence shows that the student has a good grasp of the concepts, but he misses to note a fine grain level of his knowledge about the concept.

\subsection{Difficulties which exhibit content error}
Among problems which require transitions, three difficulties are classified as content error only (Table \ref{tab:2}). 

For Cs2, \textit{tendency to treat time-dependent phase factors as decaying exponentials}, the authors\cite{Emigh2015} provided a student's reasoning:
\begin{description}
\item[Student] Since the wave equation will gain a $e^\frac{-E_{2}t}{\hbar}$ term to represent its evolution as time goes on, the probability of finding the particle in the marked area will decrease [$\dots$] since the square of its wave equation will decrease as well. 
\end{description} 

This student is in the same frame as that promoted by the problem-statement (Figure \ref{fig:Emighgraph}), which is conceptual physics, as discussed in section $\ref{sec:Ex1}$. The only difference is that the student's response is with regard to the stationary wave function. %\ecs{which is\dots}. 
The student has determined the energy of the stationary state $E_{2}$ and knows how to perform the appropriate calculations to find the probability density. However, their exponential term has a (negative) real power instead of an imaginary one. We interpret this as a content error: the student has activated incorrect resources and reasoned from them. 

\begin{dtable}{Difficulties that exhibit only content error in problems which require shifting frames. \label{tab:2} These difficulties are labeled ``C'' for content errors; and ``s'' because their problems require shifting. The ordering of the difficulties in the table is for labeling purposes only and does not represent a hierarchy.} 	
%Cs1 & Inability to differentiate between $e^\frac{-iHt}{\hbar}$ and $e^\frac{-iEt}{\hbar}$ & \cite{Singh2015}\\
\hline
Cs1 & Tendency to treat wave functions for bound systems as traveling waves  & \cite{Emigh2015}\\
\hline
Cs2 & Tendency to treat time-dependent phase factors as decaying exponentials & \cite{Emigh2015}\\
\hline
Cs3 & Difficulties related to outside knowledge in student understanding of energy measurements& \cite{Passante2015}\\
\end{dtable}

In simpler problems that do not require transition, we found an additional nine difficulties (Table \ref{tab:8}), which exhibit content error.
% * <jeremy79@ksu.edu> 2018-11-27T15:40:25.729Z:
% 
% Recommend that you move Table V to directly under Table IV. It's current place with Table VI does not make sense in light of the extended discussion of its contents at this point in the body text. 
% 
% ^.
The difficulty, \textit{incorrect belief that the time evolution of a wave function is always via an overall phase factor of the type $e^\frac{-iEt}{\hbar}$} (Cn1) in table \ref{tab:8} shows that students are performing a content error.
The problem (Figure \ref{fig:Singhgraph}) asks students to find the time dependent wave function $\Psi(x,t)$ for a system in an initial state of superpositions of the ground state and first excited states, $\Psi(x,t=0)=$$\sqrt\frac{2}{7}\phi_{1}({x})+\sqrt\frac{5}{7}\phi_{2}({x})$. The equations of the eigenfunctions and the eigenvalues are given in the problem statement. The frame of this problem is algorithmic physics. The problem statement asks the student to write down the wave function as opposed to figure out the wave function. The frame of the problem requires the student to recall the time phase factor and follow simple algorithmic steps to assign the readout energy eigenvalues from the problem statement into the time phase factor for each term and write the time dependent wave function in terms of $\phi_{1}$ and $\phi_{2}$. 
About one third of the students ($N = 202$) in this study wrote:

\begin{description}
\item[Student]  $\Psi(x, t)=\psi(x, 0)e^\frac{-iEt}{\hbar}$
\end{description} 

 The frame of the student is algorithmic physics, which is the frame of the problem. The student has written down a time phase factor, but still needs to include more content and take more steps.
This student does not read the information regarding the energy eigenvalues from the given equations in the problem statement, and does not attempt to write the answer in terms of $\phi_{1}$ and $\phi_{2}$. Therefore, this difficulty is categorized as a content error as the student is not reading enough content from the problem statement.

The difficulty, \textit{inability to differentiate between $e^\frac{-iHt}{\hbar}$ and $e^\frac{-iEt}{\hbar}$} (Cn9, Table \ref{tab:8}), occurs when students misapply the energy eigenstate instead of the Hamiltonian operator in the time evolution operator \cite{Singh2015}. The problem asks students to find the time dependent wave function $\Psi(x,t)$ for a system in an initial state of superpositions of the ground state and first excited states, $\Psi(x,t=0)=$$\sqrt\frac{2}{7}\phi_{1}({x})+\sqrt\frac{5}{7}\phi_{2}({x})$. The equations of the eigenfunctions and the eigenvalues are given in the problem statement. We frame this problem as algorithmic physics as it requires the student to recall the time phase factor and follow simple algorithmic steps: assign the readout energy eigenvalues from the problem statement into the time phase factor for each term and write the time dependent wave function in terms of $\phi_{1}$ and $\phi_{2}$. The authors mention that students write an intermediate state for $\Psi(x,t)$:
\begin{align*}
\Psi(x,t)=\Psi(x,t=0)e^{-iEt/\hbar}=&\sqrt\frac{2}{7}\phi_{1}{(x)}e^{-iE_{1}t/\hbar}\\
&+\sqrt\frac{5}{7}\phi_{2}{(x)}e^{-iE_{2}t/\hbar}
\end{align*}
%\ecs{and then what? need to see how this intermediate state relates to framing.}

Since the student proceeded from an intermediate state, we presume that the student does not attempt to re-derive the relationship between a space portion and a time portion of a wave function. This student is in the frame of the problem by reading out the energy eigenvalues ${E_{1}}$ and ${E_{2}}$ and assigning each energy into the time phase factors.
However, the intermediate step does not convey any algorithmic process or physical meaning and can not lead to the final step. 

This problem is classified as a content error, since the student uses the wrong idea that the symbol $H$ as the Hamiltonian operator and the symbol $E$ as the energy eigenvalue of the system are the same. This is evidenced by further probing by the interviewer revealed the difficulty differentiating between the Hamiltonian operator and its eigenvalue. 

%\ecs{I can't parse this.  Which frame are they in? which frame should they be in? How can we tell?}

%\BM{Comment 2: added}

The difficulty in \textit{distinguishing between three-dimensional
space and Hilbert space} (Cn8) in Table \ref{tab:8} indicates that students have difficulty differentiating vectors in real 3D space from vectors in Hilbert space, such that students may not be able to distinguish between the 3D space describing the gradient of the magnetic field in the z direction, and the 2D Hilbert space for describing a spin-$\frac{1}{2}$ particle. The question is about the Stern-Gerlach experiment: ``a beam of electrons propagates along the y direction into the page, and are in the spin state of: $\frac{(\ket{\uparrow_{z}} + \ket{\downarrow_{z}})}{\sqrt{2}}$. The beam is sent through a Stern-Gerlach apparatus (SGA) with a vertical magnetic field gradient. Sketch the electron cloud pattern that you expect to see on a distant phosphor screen in the x-z plane. Explain your reasoning.'' Due to the magnetic field gradient in the z direction, the beam of electrons will experience a force and become deviated. However, electrons due to having an intrinsic angular momentum, which is their spins, split only into two directions along the z axis and form two spots on the screen. 

The frame of this question is conceptual physics, because it encourages students to think about ``what is going on" in the physical apparatus. The problem statement requires different readouts about the direction of the magnetic field gradient, or the direction of the electron beam. 
% that will help students to connect those information with the ideas that they have about spin of an electron. 
 Students are asked to use graphical representation and  justify their reasoning. 
Only 41$\%$ of the students (N= 202) answered correctly, and the rest of the students predicted that there will be only a single spot on the screen. A typical response of a student looks like:
\begin{description}
\item[Student]  All of the electrons that come out of the SGA will be spin down with expectation value $\frac{-\hbar}{2}$ because the field gradient is in $-z$ direction.

\end{description} 

This student is thinking conceptually by reading out information about the direction of the magnetic field from the problem statement (``$-z$ direction'') and connecting that to the idea of spin$-\frac{1}{2}$ and thinking that this measurement has only one outcome and thus the expectation value is $\frac{-\hbar}{2}$. However, the student needs to more carefully read out from the problem statement that the state of the system before the measurement is in the combination of two states of spin up $\ket{\uparrow_{z}}$ and spin down $\ket{\downarrow_{z}}$, and the state of the system is not just prepared in one state of down $\ket{\downarrow_{z}}$ to stay unchanged after the measurement. This problem is categorized as a content error since the student's reasoning is missing some content that blocks a correct answer. However, the description of the student's reasoning is not enough to identify which exact piece of content is missing. It could be helpful to the students to think about what it means for a beam of electrons to be in the combination of states spin up and spin down before passing through an SGA, or to think about what it means for an electron to have an intrinsic angular momentum.

The difficulty \textit{determining the outcomes of a subsequent energy measurement} (Cn2) is mostly limited to a content error.  Students are able to use the ideas of the problem statement and operate in the frame of the question, %\ecs{this frame?},
 but are activating the wrong resources to productively and correctly solve the problem. 

%\BM{Comment 6: added}
%\ecs{I think we should take this example out and keep the one after.}
For example, a student is asked about the outcomes of an energy measurement after a previous measurement on the system of a particle in an infinite square well in an initial state $\Psi(x,0)=0.6 \Psi_{1}(x,0)+0.8i\Psi_{2}(x,0)$. Part A of the question asks ``What value or values would a measurement of the energy yield?'' Part B of the question asks what would be the result of a second energy measurement after time $t_{2}$.
 In response to part B:\cite{Passante2015}:
\begin{description}
\item[Student] The particle is described by a wave function with elements in both eigenstates. Although a measurement of energy collapses it to one, the possibility of the other still exists, so a second measurement could get the other $E$. 
\end{description}
%\BM{Comment 7: added} \ecs{I don't understand our resources analysis here.  What resources are at play? how do we know? are they in conflict with each other, or is the set incomplete, or is something else going on?}

The frame of this question is conceptual physics, which requires students to think about the idea that repeating an energy measurement does not change the state of the system.
Repeating an energy measurement only yields the same result as the first measurement, since the system is already collapsed to one of the energy eigenstates and is isolated from its surroundings. 

This student has activated several ideas about energy measurement on a physical system in a superposition wave function. In the first part of the response, the student acknowledges that a particle ``is described by a wave function with elements in both eigenstates", and also ``a measurement of energy collapses it to one". These ideas are both correct. With this being the case, it may be difficult to understand why the student arrives at the wrong answer despite seeming to have correct ideas about the system. 

This student acknowledges the fact that when a system collapses it has only one energy, but they also activate the idea that the probability of other energy, ``$E$'', ``still exists'' and associates this possibility with the second measurement on the system. 
 
The second part of this statement can be considered as correct if no measurement has been actually performed on the system (similar to the context of the problem in part A). For a system in a superposition state, if the system is prepared $n$ times in the exact same way and each time a measurement is made on the system, one can find the number of times that the energy measurement yields $E_{1}$, and the number of times that the energy measurement yields $E_{2}$. However, as soon as an energy measurement is made, the system collapses into one of the energy eigenstates and repeating the energy measurement yields the same result as the first measurement. 

 This student somehow decides that their knowledge of state collapse is not applicable here and a measurement possibly yields multiple possible energy values. 
 %For us, the key is in the student's usage of the word ``although''. 
 This student uses the word ``although'' to put these two ideas in opposition to each other. This student needs to activate more content from the information that the problem statement provides about the system before and after an energy measurement as well as repeated energy measurements to the system. 

 In the study by Singh et al. \cite{Singh2015, Singh2008} they showed that students have difficulty with the \textit{time development of the wave function after measurement of an observable} (Cn7). Students were asked about the wave function a long time after measurement of energy $E_{2}$ for an electron in an infinite square well. Some of the students stated similar responses that ``If you are talking about what happens at the instant you measure the energy, the wave function will be $\phi_{2}$, but if you wait long enough it will go back to the state before the measurement." The first part of the response suggests that the student is able to correctly relate the measured energy eigenvalue to the associated eigenstate of the system $\phi_{2}$ by activating the resource of an instant measurement. However, the student does not further investigate the idea that long after the measurement only a phase will be added to the eigenstate, which does not change the state of the system to any other combination of eigenstates; the system will not ``go back to the state before the measurement."

  \begin{dtable}{Difficulties that exhibit content error in simpler problems. \label{tab:8} These difficulties are labeled ``C'' for content errors; and ``n'' because their problems require no shifting. The ordering of the difficulties in the table is for labeling purposes only and does not represent a hierarchy.}	
  Cn1 & Incorrect belief that the time evolution of a wave function is always via an overall phase factor of the type $e^\frac{-iEt}{\hbar}$ & \cite{Singh2015}\\
\hline
Cn2 & Determining the outcomes of a subsequent energy measurement & \cite{Passante2015}\\
\hline
Cn3 & Difficulties with the possible outcomes of a measurement & \cite{Singh2015}\\
\hline
Cn4 & Failure to recognize that the time evolution of an isolated system is determined by the Schr{\"o}dinger equation: ``Decay reasoning'' & \cite{Passante2015}\\
\hline
Cn5 & Belief that the wave function will return to its initial state  & \cite{Emigh2015}\\
\hline
Cn6 & Failure to recognize that the time evolution of an isolated system is determined by the Schr{\"o}dinger equation: ``Diffusion reasoning'' & \cite{Passante2015}\\
\hline
Cn7 & Difficulties with time development of the wave function after measurement of an observable & \cite{Singh2015}\\	
\hline
Cn8 &Difficulties in distinguishing between three-dimensional space and Hilbert space & \cite{Singh2008}\\
\hline
Cn9 & Inability to differentiate between $e^\frac{-iHt}{\hbar}$ and $e^\frac{-iEt}{\hbar}$ & \cite{Singh2015}\\
 \end{dtable}

\section{Difficulties where more than one error type is possible}
For some difficulties, multiple error types are possible.  Additional details of student reasoning could resolve these ambiguities, but these details are either not gathered (survey data) or not available to us (interview data) for secondary analysis. 
 
 \begin{dtable}{Difficulties that exhibit both displacement and content error in problems which require transitions. \label{tab:4} These difficulties are labeled ``DC'' for displacement and content errors and ``s'' because their problems require shifting.  The ordering of the difficulties in the table is for labeling purposes only and does not represent a hierarchy.} 	
DCs1 & Confusion between the time dependence of probabilities of energy measurements and other quantities & \cite{Emigh2015}\\
\hline
DCs2 &Belief that the wave function will spread out over time  & \cite{Emigh2015}\\
 \end{dtable}

 The first difficulty with the \textit{confusion between the time dependence of probabilities of energy measurements and other quantities} (DCs1) in table \ref{tab:4} indicates a displacement or content error. The task asks about the time dependence aspect of energy probability measurements on a particle in a quantum mechanics harmonic oscillator system in the initial state, $\psi=\frac{i}{\sqrt{3}}\psi_{0}-\frac{\sqrt{2}}{\sqrt{3}}\psi_{1}$. 
% * <jeremy79@ksu.edu> 2018-11-27T17:10:52.291Z:
% 
% >  The first difficulty with the \textit{confusion between the time dependence of probabilities of energy measurements and other quantities} (DCs1) in table \ref{tab:4} indicates that  students could exhibit either displacement or content error.
% The table seems to indicate that the difficulty is a manifestation of both a content error and a displacement error (together). This body text seems to indicate that the difficulty is a manifestation of one or the other, and that available data prevents a determination of which. 
% 
% Also not sure about the use of the word "exhibit" here. The behavior is exhibited, but the underlying thinking/reasoning is implied. 
% 
% ^.
 A displacement error occurs when the student associates the time dependence aspect of the probabilities of energy measurements to the time independent properties of the ``probability density'' or the  ``wave function''. In this type of answer the student usually recalls some properties of the physical quantities without any justification, since they think their reasoning is correct the way it is first recalled. 

\begin{description}
\item[Student]  It [the energy probability] depends on the probability density. If it's time independent then no, if time dependent then yes.
\end{description}
  
 A content error occurs when the student is able to determine some of the features of the physical system by being in the frame intended by the problem; however, they are not considering all aspects of the problem context. As in the example mentioned in the study by Emigh et al. \cite{Emigh2015}, the student begins by stating, that ``A linear combo of stationary states is not stationary.'' This student is mindful that a superposition of eigenstates is not a stationary state. The student is also able to differentiate between the energy levels of each eigenstate and give a description of the system by stating ``The system will oscillate around $E_{0}$ and $E_{1}$''. This student activates the idea that the state of the system is in the combination of two states. This piece of reasoning leads to the activation of another piece of the idea that both energies can be available; in the student's words, the system can ``oscillate around'' two energies.
% * <jeremy79@ksu.edu> 2018-11-27T17:21:46.175Z:
% 
% > begins
% Not sure if this is the correct word to insert, but there is a word missing here.
% 
% ^.
This student should activate further resources in accordance to the problem statement which asks ``Are there times when the probability of measuring ${E_1}$ is zero and the probability of measuring ${E_0}$ is one?''
  
  \begin{dtable}{Difficulties that exhibit both displacement and transition errors in problems which require frame shifting.\label{tab:5} These difficulties are labeled ``DT'' for displacement and transition errors and ``s'' because their problems require shifting. The ordering of the difficulties in the table is for labeling purposes only and does not represent a hierarchy.} 	
DTs1 & Tendency to treat all wave functions as having a single phase & \cite{Emigh2015}\\
\hline
DTs2 & Finding the probability of an energy measurement from the wave function & \cite{Emigh2015}\\
 \end{dtable}
% * <jeremy79@ksu.edu> 2018-11-27T17:28:12.241Z:
% 
% Table placement suggestion: Place Table V with Table IV. Place Tables VII and VIII  with Table VI. This puts related tables together and closer to the body text that introduces them. 
% 
% ^.
 
  Table \ref{tab:5} indicates, that students' difficulties in \textit{tendency to treat all wave functions as having a single phase} (DTs1) can be mapped as a displacement error or a transition error. A displacement error indicates that the student has not attended to the frame of the question to blend the information effectively with the corresponding concepts in the task. This is similar to the described example in section \ref{sec:Ex2} from the Emigh et al. study \cite{Emigh2015}. The other possible error occurs when the student frames the task properly and is able to coordinate between frames; however, they fail to productively transition between frames to remove all the barriers (Section \ref{sec:Ex1}).
% * <jeremy79@ksu.edu> 2018-11-27T17:40:44.484Z:
% 
% >  Table \ref{tab:5} indicates, that students' difficulties in \textit{tendency to treat all wave functions as having a single phase} (DTs1) can be mapped as a displacement error or a transition error.
% Once again, the body text and the table text are at odds. Table VII uses both/and language. The body text about Table VII uses either/or language. From the context, I think you more mean the "or" language (That the displayed reasoning might be a manifestation of displacement error or transition error), but it's still a point of confusion. 
% 
% ^.

  \begin{dtable}{Difficulty that exhibits both content and transition errors in a problem which requires frame shifting. \label{tab:6} This difficulty is labeled ``CT'' for content and transition errors and ``s'' because its problem requires shifting.} 	
CTs1 & Tendency to misinterpret the real and imaginary components of the wave function & \cite{Emigh2015}\\ 
 \end{dtable}
 
The difficulty \textit{tendency to misinterpret the real and imaginary components of the wave function}(CTs1 in table \ref{tab:6}) shows that some students establish a conceptual discussion in a math frame to relate the real and imaginary parts of the wave function in the complex plane. However, viewing the problem as purely conceptual (e.g. in the conceptual math frame) prevents students from noting other related ideas in the problem statement. Shifting to the algorithmic math frame may help the student to recall other related facts to successfully solve the problem. Alternately, moving to the conceptual physics frame may help the student to better map the activated mathematical ideas to the problem situation. 

%Alternately, moving to the conceptual physics frame may help the students to reframe the unrelated ideas of their framed representation consistent with the goal of the problem.
 
  \begin{dtable}{Difficulties that exhibit all types of error in problems which require frame shifting.\label{tab:7} This difficulty is labeled ``DCT'' because it may involve all three types of error and ``s'' because the problem requires shifting.} 	
DCTs1 & Interpreting the meaning of expectation value &\cite{Singh2008}\\			
 \end{dtable}
% * <jeremy79@ksu.edu> 2018-11-27T17:47:49.110Z:
% 
% Tables IX and X should appear before the discussion on Error Rates begins and closer to their references in the body text. (Summary: Table V should go with Table IV. Tables VII and VIII should go with Table VI. Tables IX and X should go where Tables VII and VIII are currently located. This groups the tables by similar type and places them close to their references in the body text.) 
% 
% ^.
 
  Table \ref{tab:7} shows that students can exhibit different types of errors in \textit{interpreting the meaning of the expectation value} (DCTs1). As discussed in the Content Error Section (Section \ref{sec:Ex3}), the task required the student to start by recalling physics relations to calculate the expectation value. The student makes a content error when they activate an unstructured piece of their knowledge related to the TISE. This error is corrected by the interviewer \cite{Singh2008}.  The displacement error occurs when the student is outside of the problem statement frame (algorithmic physics), and writes down just a mathematical expression $\Bra{\Psi}{\hat{H}}\Ket{\Psi}$, which lacks blended information from the physical space.
  
An example of transition error is when the student is able to blend the physical meaning of the probability of the energy values with the related coefficients and then translates the problem into procedural steps. However, the student might leave an extra coefficient $(\frac{1}{2})$ in the final answer, $\frac{\frac{2}{7}E_{1}+\frac{5}{7}E_2}{2}$. This error can be removed by reviewing the solution and thinking purely conceptually about the quantity of the expectation value \cite{Singh2008}. For this example only the student's final answer was provided in the paper; no further narration from the student was available, which leaves uncertainty in our analysis.  
 
   \begin{dtable}{Difficulties that exhibit both displacement and content errors in simpler problems. \label{tab:10} This difficulty is labeled ``DC'' for displacement and content errors and ``n'' because the problem does not require shifting.} 	
DCn1 & Time dependence of expectation values: recognizing the special properties of stationary states-distinguishing between stationary
states and eigenstates of operators corresponding to
observables other than energy & \cite{Singh2015}\\	
 \end{dtable} 
%distinguishing between stationary states and other eigenstates, and recognizing their properties 

Table \ref{tab:10} shows students' difficulties with the calculation of time dependent expectation values in the context of Larmor precession for problems that do not require transition between frames. In this problem, the magnetic field
is along the z axis, which gives the Hamiltonian as $\hat{H}=-\gamma{B_{0}}\hat{S_{z}}$. Since the particle is initially in an eigenstate of the z component of spin angular momentum operator, the expectation value of any operator $Q$ will be time independent. 

Difficulties with \textit{recognizing the special properties of stationary states} could result in a content error, as students similar to this case state, that for a stationary state the commutation of the Hamiltonian and the operator $Q$ is nonzero, thus ``its expectation value must depend on time'' \cite{Singh2015}. 

\begin{description}
\item[Student] Since $\hat{S_{x}}$ does not commute with $\hat{H}$, its expectation value must depend on time. 
\end{description}

Although the student is able to apply Ehrenfest's theorem correctly, the student does not note that being in a stationary state changes the Hamiltonian in the time dependent phase factor from an operator ($e^\frac{-iHt}{\hbar}$) to a number ($e^\frac{-iEt}{\hbar}$), which commutes with the operator Q. 
In addition, difficulties with \textit{distinguishing between stationary states and eigenstates of operators other than energy} could result in a displacement error as students think that  ``if a system is initially in an eigenstate of $\hat{S_{x}}$, then only the expectation value of $S_{x}$ will not depend on time." \cite{Singh2015}

\section{Error rates}

\begin{figure}
 \includegraphics[width=\figwide]{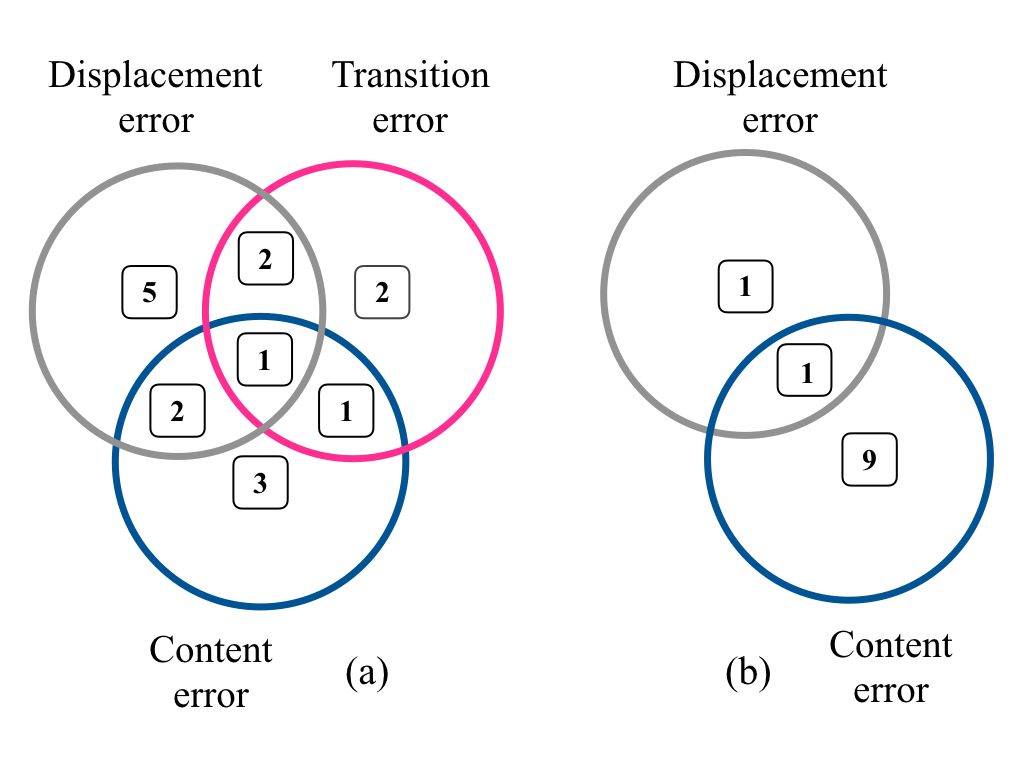}
 \caption{The number of difficulties mapped to error categories, a) for questions that require shifting, b) for questions that do not require shifting. \label{fig:4}}
% * <jeremy79@ksu.edu> 2018-11-27T22:22:31.573Z:
% 
% >  shifting
% "transition" also works, but you use "shifting" throughout the table captions
% 
% ^.
 \end{figure}

Figure \ref{fig:4} shows all the possible ways that descriptions of published difficulties can be mapped into errors in framing and resource use. Each number refers to the number of difficulties in that error category, not the number of students in that category. 

This figure shows that all the error categories and combinations of those categories are populated. %Given our set of math-in-physics frames we can predict the kinds of difficulties that will emerge for a given problem in quantum mechanics, which can provide a possible deeper structure to student's cognitive process. 

By starting with our set of math-in-physics frames and focusing on the context-dependent artifacts such as the problem statement, we can reveal a more fine-grained structure to students' cognitive processes, which are evoked in response to the keywords and cues in the problem statement.

For a problem that requires a transition, we expect to see a breakdown to displacement, content, and transition errors. If a problem does not require transition we expect to see a breakdown to displacement and content error. 
In addition, when a problem requires transition we expect to see more displacement errors because the transition inherent in the problem means students are more likely to be confused about which frames to begin with.
 
 %and c) unspecified

\begin{figure}
 \includegraphics[width=\figwide]{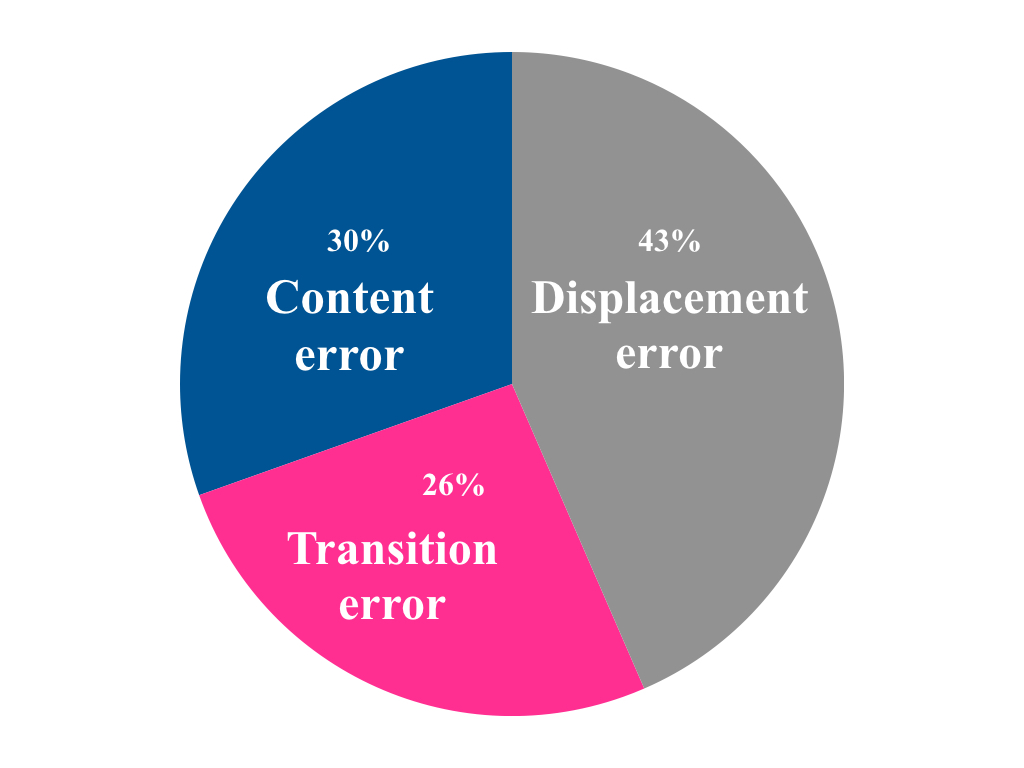}
 \caption{Displacement, transition and content error categories of difficulty topics for questions that require transition. The total percentage is not 100 due to rounding error.\label{fig:5}} 
 \end{figure}

\begin{figure}
 \includegraphics[width=\figwide]{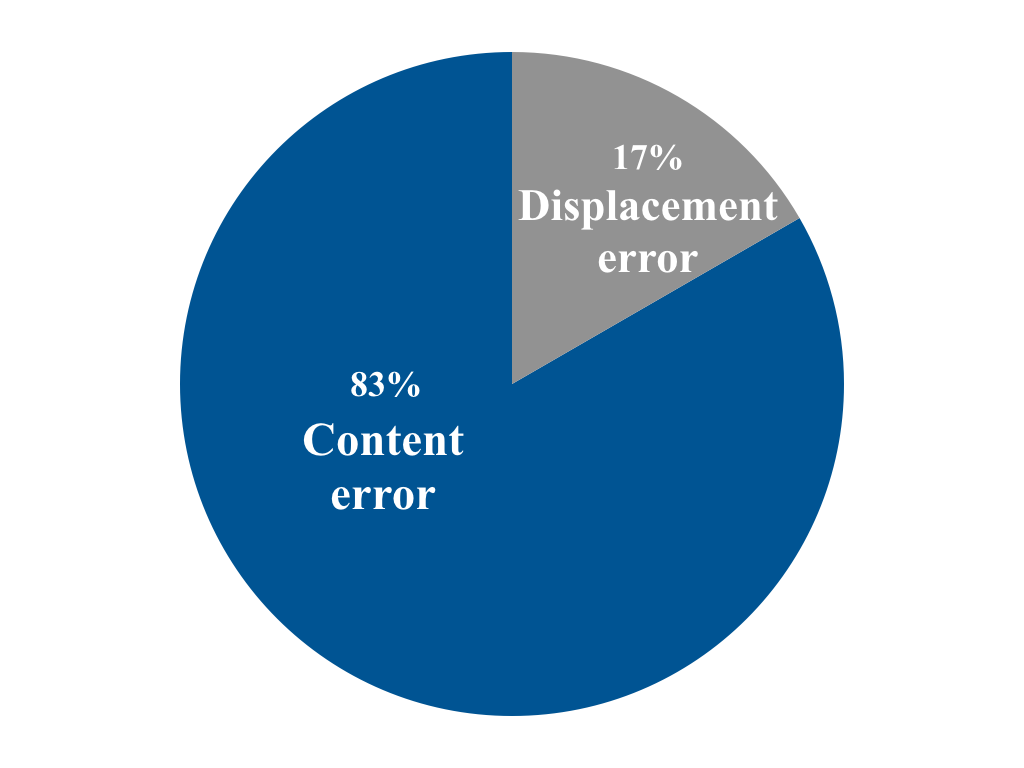}
 \caption{Displacement and content error categories of difficulty topics for questions that do not require transition\label{fig:6}} 
 \end{figure}

  Figure \ref{fig:5} and Figure \ref{fig:6} give an overview of the occurrence of the three error types -- displacement, transition, and content -- among all the topics. Figure \ref{fig:5} shows that displacement error is the most frequent among problem statements which require transition. By most frequent, we do not mean to imply that more students make that error; we mean that the displacement error category has more difficulties in it.  This distinction is important because the underlying rates of each difficulty differ in the population of students.  It could be that displacement difficulties are more common among problems that require transition because those problems are harder (overall high rate of difficulties), or that there are simply more possible ways in which students displayed regular wrong responses to those problems.
% * <jeremy79@ksu.edu> 2018-11-27T22:49:38.017Z:
% 
% It would be nice if Figure 8 and Figure 9 could be placed together, since they are so clearly related to each other. Figure 7 does not need to be a part of that group and could be moved up to page 14. 
% 
% ^.
%  This suggests that when students solve problems which require additional steps and they display a difficulty, they are more likely to begin from an inappropriate frame. In contrast, for simpler questions that do not require transition, content errors are the most frequent. On these problems, the problem statement is more likely to cue students into an appropriate frame, but students are less likely to finish successfully. 
  A great deal of effort goes into designing and testing questions which will reveal or cause student difficulties, so it's possible that these error rates are an artifact of the kinds of questions most likely to produce difficulty-like responses.
 
\section{Conclusion and Implications}

%BM-052819: added is response to reviewer comment #3 for adding a theoretical layer about the effectiveness of framing:
A difficulties view models students' conceptual understanding of fundamental ideas. Viewing knowledge as a large-grained construct makes it possible to find prevalence patterns of difficulties within a topic over a large number of students, which helps the research scale to quantitative studies. In this view, student success is determined by applying a concept to different physical situations correctly.  
 
From a knowledge in pieces view, students' ideas are fine-grained, and success is determined by how they navigate in the problem-solving space; choosing the appropriate frame, activating the relevant ideas within a frame, and transitioning out of a frame into other productive frames to coordinate different kinds of knowledge.   This emphasis on pathways in problem solving helps the research scale better to studies across different topics.

Students (similar to Eric's example and the example in the Emigh et al study \cite{Emigh2015} in section \ref{studEmighstudy}) may rely on their initial frame which may cause an error. However, students' awareness of their error due to external factors can help them to reframe the problem. This awareness can be due to an interaction with an instructor or due to a disagreement among the members of a group that arises in the context of group problem solving.  Due to the context dependent nature of ideas, the knowledge in pieces view is better able to account for students' unseen conditions and unexpected moments of  getting `stuck', aha moments and getting `unstuck’, and arriving at disagreements with an instructor or a group mate during a problem solving procedure. The knowledge in pieces framework is also better able to account for students' thought processes at a social (non-individual) level, such as students' in-class group problem solving, students working on a lab experiment, or in a tutorial session.

The goal of this paper is to reinterpret research on student difficulties in quantum mechanics through the lens of epistemological framing, as part of the family of knowledge in pieces theory, particularly using the set of four math-in-physics frames\cite{Modir2017QuantumFraming} previously applied to our observational classroom data of upper-division student problem solving. As our work in electromagnetic theory\cite{Nguyen2016StuFraming,ChariInsFraming} and quantum mechanics\cite{Modir2017QuantumFraming} has shown, the framing space is also useful in describing student ideas as they solve problems.   Additionally, these four frames can account for students' correct as well as incorrect reasoning.  
%while difficulties cannot.  

We seek an underlying structure to the kinds of difficulties that other researchers have identified, and this mission necessarily moves us away from the particular details of the topics those difficulties are tied to. Our framework splits the underlying thought processes behind student errors into three different categories as displacement error, transition error, and content error. 
Displacement error reveals a student's unproductive frame of the situation. Content error shows what pieces of knowledge have yet to be activated to understand all the ideas incorporated in the problem frame. Transition error shows that students are able to activate resources in one frame, but then cannot make a productive transition in frame to continue with problem solving. 
 We excluded nine difficulties from further analysis due to not having enough information to figure out what the framing could have been. 

Our analysis of secondary data is hampered by the very nature of that data: we do not have access to full student reasoning because sometimes primary sources do not report it, and sometimes the nature of their survey data precludes them from collecting it.  We expect that, were sufficiently detailed data available, all of the student reasoning attributed to difficulties in quantum mechanics could be analyzed using these frames.  

It is possible that if we had access to the original student data under a certain difficulty, that difficulty could possibly map to all of the framing errors, as shown in table \ref{tab:7} for one of the difficulties. We suggest that this is unlikely both statistically and theoretically.  %It could be true that, had we this data, each of these difficulties could map to all of our framing errors. 

To better illustrate this point, we use a metaphor of multidimensional space. If we attribute all twenty-seven surviving difficulties to a 27-dimensional space and the framing errors to a 3-dimensional space, we can then reproject difficulties into framing errors.   Inasmuch as our video-based data overlaps with the data presented in the difficulties papers, we notice that our data maps one difficulty to one framing error. While it is technically true that each difficulty could map to all framing errors, that interpretation is inconsistent with our video-based data and statistically implausible for the remainder of the data. 

Even though the difficulty space can be more easily scaled when adding new identified difficulty topics, the framing space is more parsimonious than the difficulties space in understanding student ideas within and across topics.  This is due to the diverse nature of knowledge in pieces approach; which supports that different students have different ways of thinking and use of their knowledge  \cite{diSessa2018}.
We suggest -- but cannot robustly support -- that many difficulties will largely map to single framing errors.  This is a potential avenue for future collaboration between difficulties-based teams and framing-based teams. At a community level, the difficulties framework provides insight into students' difficulties with specific topics and a pieces approach provides insight into dealing with students' emerging difficulties for different classroom settings in the moment, both at an individual and at a social level. 

Choosing framing over difficulties has implications for both future research and for instruction. 
For research, it is an open question as to whether these four frames -- conceptual physics, conceptual math, algorithmic physics, and algorithmic math -- constitute the optimal basis set for epistemological frames in student understanding of quantum mechanics. However, they do form a more compact basis set than is possible (let alone extant) with difficulties, as they are applicable across topics in a way which difficulties cannot be.  Because framing focuses on the pathways of problem solving, rather than students' answers and reasoning, researchers using framing generate different kinds of data and value different kinds of student responses than researchers using difficulties.
%If that were true, this work is still valuable.  

Instructors' awareness of student error categories may help them scaffold students' reasoning more effectively, as instructors can tip students into different frames\cite{Irving2013framing,ChariInsFraming} or gently nudge students to use additional resources\cite{Singh2015,Vygotsky1978} to resolve content errors. Epistemologically-aware tutorials at the introductory level\cite{ Scherr2006a} have been shown to outperform difficulties-based tutorials\cite{McDermott1998TiiP} in student understanding of Newton's Third Law\cite{Smith2007a, Sayre2012N3L}. More broadly, supporting students' epistemologies in the classroom may have far-reaching implications for retention and persistence\cite{Lising2005,Danielak2014,Geller2014a}. Curriculum development work in quantum mechanics at the upper-division is exclusively in a difficulties-based mode, though some epistemologically-aware work has occurred at the introductory level in quantum mechanics\cite{Wittmann2007}. 

Curriculum developers could take up framing as a guiding theoretical framework for professional development at the upper-division. Difficulties promotes a model of student reasoning as fractured among long lists of topically-centered wrong ideas.  A faculty member must memorize a different list of difficulties for each topic and each course, substantially increasing their overhead in teaching.  In contrast, a faculty development program using framing might emphasize a smaller set of frames to guide problem solving across topics, which may be easier for faculty to learn and apply in their teaching in different courses than long lists of difficulties.  

These differences in implications for instruction are particularly interesting for quantum mechanics because the conceptual content is epistemologically difficult\cite{Dini2017quantum} -- yet conceptually fascinating\cite{Barad1995Quantum} -- for students and because faculty who teach quantum courses usually teach other courses as well.

\begin{acknowledgments}

The authors gratefully acknowledge the contributions of the KSUPER group who participated in inter-rater reliability testing and codebook development discussions.  Beta-readers on this paper included Mary Bridget Kustusch and Matthew Sayre.  Paul Emigh pushed back hard against our characterization of difficulties as a theoretical framework, forcing us into a more productive conceptualization of this work.  An earlier version of this paper detailed both our video data analysis and the secondary analysis of difficulties, and we are grateful to the patient reviewers and  editor who advised splitting it into two papers.  Portions of this research were funded by NSF DUE-1430967, the KSU Office of Undergraduate Research and Creative Inquiry, and the KSU Physics Department.  

\end{acknowledgments}

%\bibliography{/Users/le/Dropbox/Research/Bibfiles/library}
%\bibliography{library}

\end{document}